\documentclass[11pt, a4paper]{article}
\usepackage{jheparxiv}
\usepackage[utf8]{inputenc}
\usepackage{amsmath}
\usepackage{amsfonts}
\usepackage{amssymb}
\usepackage{latexsym}
\usepackage{mathrsfs}
\usepackage{braket}		
\usepackage{graphicx}
\usepackage{color}
\usepackage{xcolor}
\usepackage{slashed}
\usepackage{twistor}
\usepackage[all]{xy}
\usepackage{tikz-cd}
\usepackage{mathtools}

\newcommand{\veps}{\varepsilon}

\newcommand{\msf}[1]{\mathsf{#1}}

\title{Celestial operator products from the worldsheet}

\author[a]{Tim Adamo,}
\author[a]{Wei Bu,}
\author[b]{Eduardo Casali,}
\author[c]{Atul Sharma}

\affiliation[a]{School of Mathematics and Maxwell Institute for Mathematical Sciences\\
University of Edinburgh, EH9 3FD, UK}
\affiliation[b]{Center for the Fundamental Laws of Nature\\ Harvard University, Cambridge, MA, 02138, USA}
\affiliation[c]{The Mathematical Institute\\
University of Oxford, Woodstock Road, OX2 6GG, UK}

\emailAdd{t.adamo@ed.ac.uk}
\emailAdd{w.bu@sms.ed.ac.uk}
\emailAdd{ecasali@g.harvard.edu}
\emailAdd{atul.sharma@maths.ox.ac.uk}

\abstract{We compute the operator product expansions of gluons and gravitons in celestial CFT from the worldsheet OPE of vertex operators of four-dimensional ambitwistor string theories. Remarkably, the worldsheet OPE localizes on the short-distance singularity between vertex operator insertions which in turn coincides with the OPE limit of operator insertions on the celestial sphere. The worldsheet CFT dynamically produces known celestial OPE coefficients -- as well as infinite towers of SL$(2,\mathbb{R})$ descendant contributions to the celestial OPE -- without any truncations or approximations. We obtain these results for all helicities and incoming/outgoing configurations. Furthermore, the worldsheet OPE encodes the infinite-dimensional symmetry algebras associated with the conformally soft sectors of gauge theory and gravity. We provide explicit operator realizations of the currents generating these symmetries on ambitwistor space in terms of vertex operators for soft gluons and gravitons, also computing their actions on hard particles of all helicities. Lastly, we show that the worldsheet OPE for momentum eigenstates produces the collinear splitting functions of gluons and gravitons.}

\begin{document} 
\maketitle
\flushbottom

\section{Introduction}
\label{Intro}


When computing scattering amplitudes, one usually expresses asymptotic states in terms of a momentum eigenstate basis, where basic physical notions (e.g., soft or collinear limits, high-energy regimes) correspond to intuitively obvious regions of the kinematic parameter space of the amplitude. However, there is nothing to prevent us from expressing amplitudes in any other basis of solutions for the asymptotic states. One particularly interesting alternative is the conformal primary basis, where massless free fields are parametrized by an arbitrary conformal dimension and a location on the celestial sphere of null directions~\cite{Pasterski:2016qvg,Pasterski:2017kqt,Pasterski:2017ylz}. In this basis, free fields transform like conformal primaries under the Lorentz group and amplitudes themselves behave like conformal correlators on the celestial sphere. Thus, scattering amplitudes evaluated in the conformal primary basis are often called `celestial amplitudes.' This raises the tantalizing possibility that there might actually exist a dynamical conformal field theory (CFT) on the celestial sphere which serves as the holographic dual to quantum field theories in (asymptotically) flat space-time.

There are many things which can be deduced about such hypothetical \emph{celestial CFTs} (CCFTs) simply by using asymptotic symmetries or known properties of scattering amplitudes in momentum space, since the conformal primary basis is related to the usual momentum eigenstate representation by a Mellin transform. Of particular interest are any constraints which can be placed on the CFT data of the celestial theory -- that is, its operator spectrum and list of operator product expansion (OPE) coefficients -- as these could help to give a realisation of CCFT and confirm celestial holography as a dynamical statement in its own right. One of the most interesting results in this regard is the determination of a class of (tree-level) OPE coefficients between CCFT primary operators representing conformal gravitons and gluons (as well as other massless fields) by Mellin transforming collinear limits of momentum space scattering amplitudes or imposing symmetry constraints on the OPE~\cite{Fan:2019emx,Pate:2019lpp,Banerjee:2020kaa,Fotopoulos:2020bqj,Banerjee:2020zlg,Banerjee:2020vnt,Guevara:2021abz,Jiang:2021xzy,Himwich:2021dau,Jiang:2021ovh}.

This OPE data has already proven a key tool in celestial holography, leading to the identification of infinite-dimensional symmetry algebras~\cite{Guevara:2021abz,Strominger:2021lvk,Himwich:2021dau,Jiang:2021ovh,Adamo:2021lrv} associated with the `conformally soft' sector\footnote{`Conformally soft' means gluons or gravitons in the conformal primary basis whose scaling dimension is such that they exhibit some factorization from the tree-level S-matrix, like gluons or gravitons with a soft momentum.}~\cite{Donnay:2018neh,Adamo:2019ipt,Puhm:2019zbl,Guevara:2019ypd} of scattering amplitudes in the conformal basis. One can also use the OPE to construct toy models and study other interesting analytic properties of celestial amplitudes, such as factorization at arbitrary numbers of points~\cite{Ebert:2020nqf}. One can also, in principle, use this to derive conformal partial wave and conformal block expansions for 4-point amplitudes, thereby probing the unitarity structure of the CCFT~\cite{Nandan:2019jas,Fan:2021isc,Atanasov:2021cje,Fan:2021pbp,Guevara:2021tvr}.

\medskip

With celestial OPE data -- as with virtually all results in celestial holography -- the question remains: is it possible to generate these results \emph{dynamically}? In this paper, we provide a positive answer for tree-level celestial OPE coefficients involving gluons and gravitons in four space-time dimensions. The key tool here is a class of chiral, constrained string theories known as \emph{ambitwistor strings}~\cite{Mason:2013sva,Geyer:2014fka}, which provide worldsheet descriptions of massless QFTs. Ambitwistor strings produce QFT scattering amplitudes from their worldsheet correlation functions in representations where all worldsheet moduli integrals are localized by constraints known as the scattering equations~\cite{Cachazo:2013hca,Cachazo:2014xea}.

The basic idea of this paper is to produce celestial OPEs using the worldsheet OPE between vertex operators representing gluons or gravitons in a conformal primary basis. Initially, it might seem obvious that this will give the correct results, since the ambitwistor string gives the right scattering amplitudes (in the conformal primary basis as well as in momentum space~\cite{Adamo:2019ipt,Casali:2020uvr}). However, this is actually far from obvious. The scattering equations tie the factorization structure of tree-amplitudes (i.e., an internal propagator going on-shell) to the structure of the boundary of the worldsheet moduli space~\cite{Cachazo:2013hca,Dolan:2013isa}, but these equations only emerge after performing the path integral for a full correlation function in the worldsheet CFT of the ambitwistor string. Our calculations only involve two `bare' vertex operators: we consider the full worldsheet OPE between these vertex operators, without reference to any larger correlator or the rest of the path integral, so the full scattering equations never actually appear in the computation.

Hence, it is not \emph{a priori} clear that the worldsheet OPE should localize to the boundary of the moduli space (i.e., where worldsheet insertion points coincide) or have anything to do with the OPE limit of operator insertions on the celestial sphere. Remarkably, this is exactly what happens: we find that the limit where worldsheet insertion points collide is directly related to the collision of insertion points on the celestial sphere; that the worldsheet OPE localizes onto these configurations; and that the expression for the worldsheet OPE exactly reproduces the structure of the OPE in CCFT. 

More specifically, the ambitwistor worldsheet OPE gives the tree-level celestial OPEs of gluons and gravitons in pure Yang-Mills theory, general relativity and Einstein-Yang-Mills theory for every helicity and incoming/outgoing configuration, complete with a full chiral tower of $\SL(2,\R)$ descendants\footnote{The only known celestial OPE coefficients we cannot reproduce involve two gluons fusing into a graviton: this is related to a well-known issue in ambitwistor strings which we will elaborate on later.}. In every case we reproduce the known results in the celestial holography literature, and in several cases extend them to include descendant contributions which have not appeared explicitly before (e.g., the $\SL(2,\R)$ descendant contributions to the celestial OPEs between incoming and outgoing gluons or gravitons). We also show that if one instead considers momentum eigenstate vertex operators, the worldsheet reproduces the tree-level collinear splitting functions of gluons or gravitons.

To appreciate how surprising these results are, one need only contrast with the analogous calculation in ordinary string theory~\cite{Jiang:2021csc}, where the worldsheet OPE between includes an infinite tower of poles (through the Koba-Nielsen factors) which does not localize anywhere in the moduli space.  Other proposals for two-dimensional theories that reproduce CCFT OPEs were put forward in~\cite{Magnea:2021fvy,Kalyanapuram:2021bvf}, but these don't seem to have a relation to the ambitwistor string we use in this paper. Moreover, these models are limited to the soft sector only, while computations with the ambitwistor string are applicable to both hard and soft operators.


The worldsheet OPE is also a powerful tool to investigate the action of holographic symmetry algebras associated with the conformally soft gluon and graviton sectors. This can be done independently (i.e., without taking soft limits of the results for generic states) by defining ambitwistor string vertex operators for conformally soft gluons and gravitons. Taking the worldsheet OPEs between these immediately produces the infinite dimensional algebras associated with these soft sectors on the celestial sphere. Furthermore, the OPEs between soft and hard vertex operators give the action of the soft sector on generic gluons and gravitons in CCFT in a novel and simple factorized form (see equations \eqref{ggfact}, \eqref{grgrfac} and \eqref{sgravhglu} in Section~\ref{Sec:Sym}).

\medskip

The paper is organized as follows. Section~\ref{Prelim} begins with a brief review of the main players in this story: the OPE in CCFT and the four-dimensional ambitwistor string. Section~\ref{same_h} then computes the worldsheet OPEs between gluons or gravitons of the same helicity, recovering their celestial OPEs. Section~\ref{Sec:Mixed} computes the worldsheet OPEs between gluons or gravitons of mixed helicity, where a new subtlety emerges. Here, the worldsheet OPE localizes on a region of the moduli space that requires a rescaled coordinate patch, but this is easily dealt with and the correct celestial OPEs are again recovered. In Section~\ref{Sec:Sym} we show that the ambitwistor worldsheet OPE also captures the structure of certain infinite-dimensional Kac-Moody and $w_{1+\infty}$-algebras associated with the conformally soft gluon and graviton sectors, respectively. Finally, Section~\ref{Sec:MomEig} outlines how a momentum eigenstate version of these calculations produces collinear splitting functions from the worldsheet OPE.


\section{Preliminaries}\label{Prelim}

In this section, we review the two main building blocks for this paper: the OPE structure of celestial conformal field theories (CCFTs) and four-dimensional ambitwistor strings. Readers who are already familiar with these tools can safely skip to Section~\ref{same_h}.


\subsection{Operator products in CCFTs}

The OPE is a familiar universal feature of CFTs, with the coefficients in this expansion forming a crucial part of the CFT data itself (along with the operator spectrum), and \emph{celestial} CFTs (CCFTs) are no exception. Determining the OPE structure of a CCFT means probing the behaviour of operator insertions on the celestial sphere when these insertions get close to each other. Fortunately, it is easy to see that this OPE limit in CCFT is closely related to the collinear limit of scattering amplitudes in a momentum eigenstate basis, which is very well-studied.

This fact follows trivially upon parametrizing outgoing null momenta $k^2=0$ as 
\be\label{csparam}
k^{\mu}=\frac{\omega}{\sqrt{2}}\left(1+|z^2|,\,z+\bar{z},\,-\im (z-\bar{z}),\,1-|z|^2\right)\,, 
\ee
where $(z,\bar{z})$ are local stereographic coordinates on the celestial sphere and $\omega$ is the energy or frequency. In CCFT, the energy $\omega$ is traded for a scaling dimension $\Delta$ by a Mellin transform; in terms of the external states of a $n$-point, on-shell scattering amplitude
\be\label{MellinTrans}
A_{n}(\{\Delta,z,\bar{z}\})=\int_{0}^{\infty}\prod_{i=1}^{n}\d \omega_i\,\omega_{i}^{\Delta_i-1}\,A_{n}(\{\omega,z,\bar{z}\})\,,
\ee
with $A_{n}(\{\omega,z,\bar{z}\})$ the momentum-space amplitude in parametrization \eqref{csparam} and 
$A_{n}(\{\Delta,z,\bar{z}\})$ the \emph{celestial amplitude}. This Mellin transform ensures that massless particles are para-- metrized by a position on the celestial sphere and a scaling dimension, that their wavefunctions transform naturally under the conformal group and that their scattering amplitudes transform like 2d conformal correlators with the appropriate scaling dimensions~\cite{Pasterski:2016qvg,Pasterski:2017ylz,Pasterski:2017kqt}. The inner product of any two null momenta parametrized in this way obeys $k_i\cdot k_{j}\propto|z_i-z_j|^2$, so the momenta become collinear as their positions on the celestial sphere coincide $(z_i,\bar{z}_i)\to(z_j,\bar{z}_j)$, and \emph{vice versa}.

Supposing the existence of a dynamical CCFT which manifests these properties, let $\cO_{i}(z_i,\bar{z}_i)$ denote the local primary operator on the celestial sphere corresponding to a massless physical state of conformal dimension $\Delta_i$; \eqref{csparam} captures the relationship between the operator insertion point and the on-shell 4-momentum of the corresponding massless particle in the asymptotically flat bulk. Conformal symmetry fixes the OPE between such conformal primaries to 
\begin{equation}\label{genOPE}
    \cO_i(z,\bar{z})\,\cO_j(0) = \sum_{k\in I} C_{ijk}(z,\bar{z},\p,\dbar)\,\cO_k(0)\,,
\end{equation}
$I$ is the set of admissible scaling dimensions and $C_{ijk}$ are the OPE coefficients which are singular in $(z,\bar{z})$ and depend on $\Delta_i$, $\Delta_j$ and $\Delta_k$ as well as the corresponding operator spins.


If $G(1,2,\ldots,n)$ denotes a correlator of $n$ primary operators, the OPE expansion can be used in the limit $(z_{1},\bar{z}_1)\rightarrow (z_2,\bar{z}_2)$ to write 
\begin{equation}
    G(1,2,\dots,n)\overset{(z_1,\bar{z}_1)\sim (z_2,\bar{z}_2)}{=}\sum_{k\in I}C_{ijk}(z_{12},\bar{z}_{12},\p_2,\dbar_2)\,\langle\cO_k(z_2,\bar{z}_2)\cO_3(z_3,\bar{z}_3)\cdots\cO_n(z_n,\bar{z}_n)\rangle\,,
\end{equation}
where $z_{12}:=z_1-z_2$ and $\bar{z}_{12}:=\bar{z}_1-\bar{z}_2$. Thus, OPE data can be read off from the short distance behaviour of correlation functions. In the context of CCFT, where correlation functions on the celestial sphere are related to usual momentum space scattering amplitudes by a Mellin transform, this means that OPE data can be read off by Mellin-transforming the collinear behaviour of the amplitude in momentum space~\cite{Fan:2019emx}. 

If $(z,\bar{z})$ are independent variables, one can separately consider holomorphic and anti-holomorphic OPE limits where $z_{ij}\to 0$ and $\bar{z}_{ij}\to0$. In momentum space, where a massless 4-momentum admits a spinor decomposition $k^{\alpha\dot\alpha}=\kappa^{\alpha}\,\tilde{\kappa}^{\dot\alpha}$, this corresponds to $\kappa^{\alpha}$ and $\tilde{\kappa}^{\dot\alpha}$ being independent, and the separate OPE limits correspond to chiral collinear limits where 
\begin{equation}
\epsilon_{\beta\alpha}\,\kappa_{i}^{\alpha}\,\kappa_{j}^{\beta}:=\la i\,j\ra\to0\,, \quad \mbox{ or } \quad \epsilon_{\dot\beta\dot\alpha}\,\tilde{\kappa}_{i}^{\dot\alpha}\,\tilde{\kappa}_{j}^{\dot\beta}:=[i\,j]\to0\,,
\end{equation}
independently. For Lorentzian real kinematics, or on the celestial sphere of null directions in real Minkowski space, these limits cannot be taken independently. However, by passing to $(2,2)$ signature (where the celestial sphere becomes the celestial torus)~\cite{Atanasov:2021oyu} or complexifying the celestial sphere, one does have truly independent holomorphic and anti-holomorphic limits. Throughout this paper, we assume that the OPE limits can be taken independently, although the precise way in which this is realized is not important. Thus, from now on we assume that $(z,\bar{z})$ are independent variables, and if desired one can simply read `celestial sphere' as `celestial torus' -- the only difference is whether $(z,\bar{z})$ are viewed as complex or real, respectively.

The relationship between collinear limits and CCFT OPE limits was used to calculate the leading contributions to the (tree-level) OPEs among celestial gluon and graviton operators~\cite{Fan:2019emx,Pate:2019lpp} providing the first concrete OPE data for CCFTs. Let $\cO^{\mathsf{a}}_{\pm,\Delta}(z,\bar{z})$ denote the CCFT primary operator corresponding to an outgoing positive/negative helicity gluon inserted at $(z,\bar{z})$ on the celestial sphere/torus with scaling dimension $\Delta$. The index $\mathsf{a}$ is valued in the adjoint of the gauge group. Then the leading OPE between gluons of the same helicity is~\cite{Fan:2019emx,Pate:2019lpp}\footnote{Our structure constants are normalised to remove overall factors of $-\im$.}
\be\label{shgluon0}
\cO^{\msf{a}}_{+,\Delta_i}(z_i,\bar{z}_i)\, \cO^{\msf{b}}_{+,\Delta_j}(z_j,\bar{z}_j)\sim\frac{f^{\msf{abc}}}{z_{ij}}\,B(\Delta_i-1,\Delta_j-1)\,\cO^{\msf{c}}_{+,\Delta_i+\Delta_j-1}(z_j,\bar{z}_j)\,
\ee
where $f^{\msf{abc}}$ are the structure constants of the gauge group and $B(x,y)$ denotes the Euler Beta function
\be\label{betafunction}
B(x,y):=\frac{\Gamma(x)\,\Gamma(y)}{\Gamma(x+y)}=\int_{0}^{\infty}\frac{\d t\,t^{x-1}}{(1+t)^{x+y}}=\int_{0}^{1}\d t\,t^{x-1}\,(1-t)^{y-1}\,.
\ee
Here we have included two useful integral representations of the Beta function. The leading OPE between outgoing gluons of opposite helicity includes both holomorphic and anti-holomorphic singularities:
\begin{multline}\label{ohgluon0}
\cO^{\msf{a}}_{+,\Delta_i}(z_i,\bar{z}_i)\, \cO^{\msf{b}}_{-,\Delta_j}(z_j,\bar{z}_j)\sim\frac{f^{\msf{abc}}}{z_{ij}}\,B(\Delta_i-1,\Delta_j+1)\,\cO^{\msf{c}}_{-,\Delta_i+\Delta_j-1}(z_j,\bar{z}_j) \\
 +\frac{f^{\msf{abc}}}{\bar{z}_{ij}}\,B(\Delta_i+1,\Delta_j-1)\,\cO^{\msf{c}}_{+,\Delta_i+\Delta_j-1}(z_j,\bar{z}_j)\,,
\end{multline}
where additional terms which can arise by coupling to gravity have been omitted.

Following this strategy of Mellin transforming the collinear splitting functions in momentum space, it is straightforward to write down similar leading OPEs between gravitons and gravitons or gravitons and gluons, as well as all incoming/outgoing configurations~\cite{Fan:2019emx,Pate:2019lpp}. It is also possible to derive or generalise these results with asymptotic symmetry constraints~\cite{Pate:2019lpp,Banerjee:2020kaa,Himwich:2021dau,Jiang:2021ovh,Ebert:2020nqf}. Using BCFW recursion and conformal blocks one can determine the full family of holomorphic or anti-holomorphic $\SL(2,\R)$-descendant contributions to the OPE with a fixed holomorphic/anti-holomorphic singularity~\cite{Guevara:2021abz,Himwich:2021dau,Jiang:2021ovh}.

Thus, the known CCFT OPE data can be stated in terms of these holomorphic/anti-holomorphic $\SL(2,\R)$ descendant families. For outgoing gluons, the OPEs \eqref{shgluon0} and \eqref{ohgluon0} are extended to  
\be\label{shgluon}
\cO^{\msf{a}}_{+,\Delta_i}\, \cO^{\msf{b}}_{+,\Delta_j}\sim\frac{f^{\msf{abc}}}{z_{ij}}\,\sum_{m=0}^{\infty}B(\Delta_i+m-1,\Delta_j-1)\,\frac{\bar{z}_{ij}^{m}}{m!}\,\dbar^{m}_{j}\cO^{\msf{c}}_{+,\Delta_i+\Delta_j-1}(z_j,\bar{z}_j)\,,
\ee
and
\begin{multline}\label{ohgluon}
\cO^{\msf{a}}_{+,\Delta_i}\, \cO^{\msf{b}}_{-,\Delta_j}\sim\frac{f^{\msf{abc}}}{z_{ij}}\,\sum_{m=0}^{\infty}B(\Delta_i+m-1,\Delta_j+1)\,\frac{\bar{z}_{ij}^{m}}{m!}\,\dbar^{m}_j\cO^{\msf{c}}_{-,\Delta_i+\Delta_j-1}(z_j,\bar{z}_j) \\
 +\frac{f^{\msf{abc}}}{\bar{z}_{ji}}\,\sum_{m=0}^{\infty}B(\Delta_i+1,\Delta_j+m-1)\,\frac{z_{ji}^m}{m!}\,\partial_i^m\cO^{\msf{c}}_{+,\Delta_i+\Delta_j-1}(z_i,\bar{z}_i)\,,
\end{multline}
where $\partial_i\equiv\frac{\partial}{\partial z_i}$, $\dbar_j\equiv\frac{\partial}{\partial\bar{z}_j}$ and the obvious insertion-point dependence of the operators on the left-hand-side of the OPE has been suppressed. Analogous results can also be obtained for OPEs involving incoming and outgoing states as well as gravitons; rather than recapitulating all of the relevant formulae here we refer the reader to the literature~\cite{Pate:2019lpp,Banerjee:2020kaa,Himwich:2021dau,Jiang:2021ovh} or to the following sections where they are derived. 

We will be interested in CCFTs corresponding to tree-level pure Yang-Mills, gravity and Einstein-Yang-Mills (EYM), although it is also possible to give explicit formulae for gauge theories and gravities with higher-dimensional operators or generic cubic couplings~\cite{Pate:2019lpp,Himwich:2021dau,Jiang:2021ovh}.


\subsection{Ambitwistor strings in four space-time dimensions}

Ambitwistor strings are two-dimensional CFTs governing holomorphic maps from closed Riemann surfaces to \emph{ambitwistor space}~\cite{Mason:2013sva,Berkovits:2013xba,Adamo:2013tsa}, the space of complexified null geodesics (up to scale) in any space-time~\cite{Lebrun:1983pa}. As these are the central tool for the remainder of this paper, we provide a brief overview of ambitwistor strings here.

In four-dimensional Minkowski space-time, ambitwistor space can be parametrized by twistor and dual twistor variables. Let $Z^{A}=(\mu^{\dot\alpha},\lambda_{\alpha})$ denote homogeneous coordinates on $\CP^3$, and define twistor space $\PT=\{Z\in\CP^3|\lambda_{\alpha}\neq0\}$ to be the open subset of $\CP^3$ with the projective line $\{\lambda_\alpha=0\}$ removed. Likewise, take $W_{A}=(\tilde{\lambda}_{\dot\alpha},\tilde{\mu}^{\alpha})$ to be homogeneous coordinates on the projective dual twistor space $\PT^*$. Ambitwistor space is then the projective quadric
\be\label{PA}
\PA=\{(Z,W)\in\PT\times\PT^*\,|\,Z\cdot W=0\}\,, \qquad \mbox{where }\: Z\cdot W=[\mu\,\tilde{\lambda}]+\la\tilde{\mu}\,\lambda\ra\,,
\ee
making use of the usual spinor-helicity notation $[\mu\,\tilde{\lambda}]:=\mu^{\dot\alpha}\tilde{\lambda}_{\dot\alpha}$, $\la\tilde{\mu}\,\lambda\ra:=\tilde\mu^{\alpha}\lambda_{\alpha}$.

Ambitwistor space is actually equivalent to the cotangent bundle of complexified null infinity up to a scale: $\PA\cong\P(T^*\scri_\C)$~\cite{Geyer:2014lca}. In particular, the five complex dimensions of $\PA$ correspond to a choice of point on $\scri_\C$ and complexified null direction, which selects a complexified null geodesic in Minkowski space up to scale as required. More generally, $\PA$ is related non-locally to space-time: a point in $\PA$ is a complexified null geodesic in space-time and a point in Minkowski space corresponds to the projective quadric $\CP^1\times\CP^1$ in $\PA$, representing the space of complexified null directions through that point. 

\medskip

The 4-dimensional ambitwistor string takes $Z^A$, $W_{A}$ as its main worldsheet variables, assigning bosonic statistics and conformal weight $(\frac{1}{2},0)$ to each; in other words they are bosonic worldsheet spinors $Z^{A}, W_{A}\in\Omega^0_{\Sigma}(K^{1/2}\otimes\C^4)$, where $\Sigma$ is a closed Riemann surface and $K$ is the canonical bundle of $\Sigma$. The prototypical worldsheet action takes the form~\cite{Geyer:2014fka}
\be\label{ATS}
S=\frac{1}{2\,\pi}\int_{\Sigma}W_{A}\,\dbar Z^{A}-Z^{A}\,\dbar W_A+e\,Z\cdot W\,,
\ee
where $\dbar$ is the anti-holomorphic Dolbeault operator on $\Sigma$ and $e\in\Omega^{0,1}_{\Sigma}$ is a worldsheet GL$(1,\C)$ gauge field that serves as a Lagrange multiplier to enforce the quadric constraint $Z\cdot W=0$ and restrict the target space\footnote{Projective scalings of $Z^A$ and $W_A$ arise in the worldsheet theory from the fact that, as sections of $K^{1/2}$, they do not come with a fixed scale on $\Sigma$.} to $\PA$. The model \eqref{ATS} has holomorphic reparametrization invariance as well as gauge transformations associated to the $\GL(1,\C)$ gauge field $e$ which must be fixed (e.g., with the usual BRST procedure) in order to quantize the theory.

The basic action \eqref{ATS} can be supplemented with additional worldsheet matter to obtain physically interesting theories. For instance, if a worldsheet current algebra (characterised by the rank of the gauge group and a level $k\in\N$) is added, then the theory can be gauge-fixed and quantised; this model is often referred to as the heterotic ambitwistor string.\footnote{With appropriate target-space supersymmetry and rank and level of the current algebra, this worldsheet CFT is truly anomaly-free, but these anomalies actually have little effect on the local OPE calculations of interest in this paper; we only require the vertex operators from the BRST cohomology.} The BRST cohomology contains vertex operators of the form
\be\label{GluonVOs}
\cU^{\msf{a}}_{+}=\int_{\Sigma}j^{\msf{a}}\,a(Z)\,, \qquad \cU^{\msf{a}}_{-}=\int_{\Sigma} j^{\msf{a}}\,\tilde{a}(W)\,,
\ee
where $j^{\msf{a}}$ is the worldsheet current of the gauge group with conformal weight $(1,0)$ and $a(Z)$, $\tilde{a}(W)$ are positive and negative helicity gluon wavefunctions on twistor and dual twistor space, respectively. These wavefunctions are cohomology classes of homogeneity degree zero: $a\in H^{0,1}(\PT,\cO)$ and $\tilde{a}\in H^{0,1}(\PT^*,\cO)$ which in turn define positive and negative helicity spin-1 free fields on Minkowski space through the usual Penrose transform integral formulae:
\be\label{PTrans}
\begin{split}
\tilde{F}_{\dot\alpha\dot\beta}(x)&=\int \la\lambda\,\d\lambda\ra\,\left.\frac{\partial^2 a}{\partial\mu^{\dot\alpha}\partial\mu^{\dot\beta}}\right|_{\mu^{\dot\alpha}=x^{\alpha\dot\alpha}\lambda_\alpha}\, \\
 F_{\alpha\beta}(x)&=\int [\tilde{\lambda}\,\d\tilde{\lambda}]\,\left.\frac{\partial^2 \tilde{a}}{\partial\tilde{\mu}^{\alpha}\partial\tilde{\mu}^{\beta}}\right|_{\tilde{\mu}^{\alpha}=-x^{\alpha\dot\alpha}\tilde{\lambda}_{\dot\alpha}}\,.
\end{split}
\ee
The particular representation of the free field (e.g., momentum eigenstate or boost eigenstate) depends on the choice of cohomology representative on twistor or dual twistor space.

Correlation functions of the vertex operators \eqref{GluonVOs} -- including various ghost contributions -- in the heterotic ambitwistor string at genus zero produce the complete tree-level S-matrix of four-dimensional Yang-Mills theory~\cite{Geyer:2014fka,Geyer:2016nsh}. Crucially, there are only two worldsheet OPEs to account for in the heterotic ambitwistor string: the first is the free symplectic boson system defined by the kinetic term in \eqref{ATS} after gauge fixing $a=0$:
\be\label{symp_bosons}
 Z^A(\sigma)\,W_B(\sigma')\sim \frac{\delta^A_B\,\sqrt{\d\sigma\,\d\sigma'}}{\sigma-\sigma'}\,,
\end{equation}
where $\sigma$ is a local affine coordinate on the worldsheet $\Sigma$. The other relevant worldsheet OPE is that of the worldsheet current algebra:
\be\label{wcaOPE*}
j^{\msf{a}}(\sigma)\,j^{\msf{b}}(\sigma')\sim\frac{k\,\delta^{\msf{ab}}}{(\sigma-\sigma')^2}\,\d\sigma\,\d\sigma'+\frac{f^{\msf{abc}}\,j^{\msf{c}}(\sigma')}{\sigma-\sigma'}\,\d\sigma\,,
\ee
where $k$ is the level of the worldsheet current algebra, $\delta^{\msf{ab}}$ is the Killing form of the gauge group and $f^{\msf{abc}}$ are the structure constants. The double pole in this OPE is related to multi-trace contributions to worldsheet correlators mediated by non-unitary gravitational degrees of freedom in the heterotic ambitwistor string~\cite{Azevedo:2017lkz,Berkovits:2018jvm}. These can be systematically eliminated by sending $k\to0$; while global properties of the worldsheet current algebra require $k$ to be a positive integer, setting $k=0$ is simply a formal prescription to decouple gravitational modes at genus zero which still produces sensible worldsheet OPEs (since these are insensitive to global features of the worldsheet theory)~\cite{Berkovits:2004jj,Adamo:2018hzd}. Thus, we will assume that the worldsheet current algebra has only the simple OPE
\be\label{wcaOPE}
j^{\msf{a}}(\sigma)\,j^{\msf{b}}(\sigma')\sim\frac{f^{\msf{abc}}\,j^{\msf{c}}(\sigma')}{\sigma-\sigma'}\,\d\sigma\,,
\ee
from now on.

\medskip

The basic worldsheet action \eqref{ATS} can also be modified to describe four-dimensional Einstein gravity. Rather than adding a worldsheet current algebra, one adds a complex worldsheet fermion system~\cite{Geyer:2014fka}:
\be\label{wsferm}
\frac{1}{2\pi}\int_{\Sigma}\tilde{\rho}_{A}\,\dbar\rho^{A}+\cdots\,,
\ee
where $\tilde{\rho}_{A}$, $\rho^{A}$ have conformal weight $(\frac{1}{2},0)$ and `$+\cdots$' includes various symmetry currents which are sensitive to the conformal structure of Minkowski space and form a closed worldsheet algebra with $Z\cdot W$. After gauge-fixing and quantizing, the BRST cohomology of this ambitwistor string contains vertex operators for positive and negative helicity gravitons:
\be\label{GravVOs}
\begin{split}
\cV_{+}&=\int_{\Sigma}\left[\tilde{\lambda}\,\frac{\partial h(Z)}{\partial\mu}\right]+\tilde{\rho}^{\dot\alpha}\,\rho^{\dot\beta}\,\frac{\partial^2 h(Z)}{\partial\mu^{\dot\alpha}\partial\mu^{\dot\beta}}\,, \\
\cV_{-}&=\int_{\Sigma}\left\la\lambda\,\frac{\partial\tilde{h}(W)}{\partial\tilde{\mu}}\right\ra+\rho^{\alpha}\,\tilde{\rho}^{\beta}\,\frac{\partial^2 \tilde{h}(W)}{\partial\tilde{\mu}^{\alpha}\partial\tilde{\mu}^{\beta}}\,,
\end{split}
\ee
where $h(Z)$ and $\tilde{h}(W)$ are cohomology classes of homogeneity weight two: $h\in H^{0,1}(\PT,\cO(2))$ and $\tilde{h}\in H^{0,1}(\PT^*,\cO(2))$. As in the gauge theory case, one makes specific choices for these cohomology representatives corresponding to the desired free field representation on space-time.

After accounting for ghosts and subtleties associated with the worldsheet fermions (dealt with using the standard picture changing formalism), correlation functions of the vertex operators \eqref{GravVOs} in the genus zero ambitwistor string produce the complete tree-level S-matrix of general relativity in Minkowski space~\cite{Geyer:2014fka,Geyer:2016nsh}. In addition to the twistor/dual twistor symplectic bosons \eqref{symp_bosons}, the gravitational ambitwistor string contains only one other worldsheet matter OPE:
\be\label{wsferm}
\rho^{A}(\sigma)\,\tilde{\rho}_{B}(\sigma')\sim \frac{\delta^{A}_{B}\,\sqrt{\d\sigma\,\d\sigma'}}{\sigma-\sigma'}\,.
\ee
In practical computations, it is of course crucial that the worldsheet OPEs \eqref{symp_bosons} and \eqref{wsferm} are applied consistently with the fields' statistics (bosonic and fermionic, respectively).


\paragraph{Conformal primary vertex operators:} Gluons and gravitons in the conformal primary basis naturally adapted to CCFT~\cite{Pasterski:2017ylz,Pasterski:2017kqt} are represented in the ambitwistor string by choosing appropriate wavefunctions on twistor or dual twistor space. For gluons of conformal dimension $\Delta$ inserted at $(z,\bar{z})$ on the celestial sphere, these wavefunctions are~\cite{Adamo:2019ipt}
\be\label{cgluonsf}
\begin{split}
a(Z)&=\int\limits_{\C^*\times\R_+}\frac{\d s}{s}\,\frac{\d t}{t^{2-\Delta}}\,\bar\delta^2(z-s\,\lambda)\,\e^{\im\veps\,t\,s\,[\mu\,\bar z]}\,, \\
\tilde{a}(W)&=\int\limits_{\C^*\times\R_+}\frac{\d \tilde{s}}{\tilde{s}}\,\frac{\d t}{t^{2-\Delta}}\,\bar{\delta}^{2}(\bar{z}-\tilde{s}\,\tilde{\lambda})\,\e^{\im\veps\,t\,\tilde{s}\,\la\tilde{\mu}\,z\ra}\,,
\end{split}
\ee
where $z_{\alpha}=(1,z)$, $\bar{z}_{\dot\alpha}=(1,\bar{z})$ represent the position on the celestial sphere in terms of homogeneous coordinates on a patch of $\CP^1$, $\veps=\pm1$ labels whether the gluon is incoming ($\veps=-1$) or outgoing ($\veps=1$)\footnote{Strictly speaking, one should associate $\veps$ with a particular chirality of homogeneous coordinate on $\CP^1$ (usually $\bar{z}_{\dot\alpha}$). This leads to an asymmetry in the appearance of the wavefunctions, but for integer spin this asymmetry can always be removed by rescaling various parameters to obtain \eqref{cgluonsf}. For half-integer-spin fields, more care would be required.} and
\be\label{hdelta2}
\bar{\delta}^{2}(z-s\,\lambda):=\frac{1}{(2\pi\im)^2}\,\bigwedge_{\alpha=0,1}\dbar\left(\frac{1}{z_{\alpha}-s\,\lambda_{\alpha}}\right)\,,
\ee
is a double holomorphic delta function, with support only where both components ($\alpha=0,1$) of its argument vanish.

Each representative comes with two integrals: one is a Mellin integral running over $\R_{+}=(0,\infty)$ parametrized by $t$; the other is an affine scale integral (parametrized by $s$ or $\tilde{s}$) over all non-zero complex numbers which ensures the correct homogeneity weight on $\PT$ or $\PT^*$. These integrals can be performed explicitly -- for instance
\be
a=(-\im\,\veps)^{1-\Delta}\,\Gamma(\Delta-1)\left(\frac{\la\xi\,\lambda\ra}{\la\xi\,z\ra}\right)^{\Delta}\,\frac{\bar{\delta}(\la \lambda\,z\ra)}{[\mu\,\bar{z}]^{\Delta-1}}\,,
\ee
where
\be\label{hdelta1}
\bar{\delta}(\la\lambda\,z\ra):=\frac{1}{2\pi\im}\dbar\left(\frac{1}{\la\lambda\,z\ra}\right)\,,
\ee
is a single holomorphic delta function with support only where its argument vanishes. While this fully-integrated form manifests poles at $\Delta=1,0,-1,\ldots$ corresponding to conformally soft gluons~\cite{Donnay:2018neh,Fan:2019emx,Pate:2019mfs,Adamo:2019ipt}, it will be more useful for us to keep the integrals in play for future calculations. In particular, because the scale integral is affine, the vertex operators in this representation are not globally-defined on the ambitwistor string moduli space. By keeping the scale integrals explicit, it is easy to change variables to suit different coordinate patches on the moduli space -- this cannot be done one the integrals have been performed. The ambitwistor string vertex operators for these conformal primary representatives are denoted
\be\label{cglvo+}
\cU^{\msf{a},\veps}_{+,\Delta}(z,\bar{z})=\int\limits_{\Sigma\times\C^*\times\R_+}j^{\msf{a}}(\sigma)\;\frac{\d s}{s}\,\frac{\d t}{t^{2-\Delta}}\,\bar\delta^2(z-s\,\lambda(\sigma))\,\e^{\im\veps\,t\,s\,[\mu(\sigma)\,\bar z]}\,, 
\ee
and
\be\label{cglvo-}
\cU^{\msf{a},\veps}_{-,\Delta}(z,\bar{z})=\int\limits_{\Sigma\times\C^*\times\R_+}j^{\msf{a}}(\sigma)\;\frac{\d \tilde{s}}{\tilde{s}}\,\frac{\d t}{t^{2-\Delta}}\,\bar{\delta}^{2}(\bar{z}-\tilde{s}\,\tilde{\lambda}(\sigma))\,\e^{\im\veps\,t\,\tilde{s}\,\la\tilde{\mu}(\sigma)\,z\ra}\,.
\ee
From now on, we suppress the ranges of integration in vertex operators, trusting that this can be inferred from the structure of the integrand.

The ambitwistor representatives for positive and negative helicity gravitons in the conformal primary basis have a similar structure~\cite{Adamo:2019ipt}:
\be\label{cgravsf}
\begin{split}
h(Z)&=\int\frac{\d s}{s^3}\,\frac{\d t}{t^{3-\Delta}}\,\bar\delta^2(z-s\,\lambda)\,\e^{\im\veps\,t\,s\,[\mu\,\bar z]}\,, \\
\tilde{h}(W)&=\int\frac{\d \tilde{s}}{\tilde{s}^3}\,\frac{\d t}{t^{3-\Delta}}\,\bar{\delta}^{2}(\bar{z}-\tilde{s}\,\tilde{\lambda})\,\e^{\im\veps\,t\,\tilde{s}\,\la\tilde{\mu}\,z\ra}\,,
\end{split}
\ee
differing from \eqref{cgluonsf} in the powers of the scaling parameters which account for the fact that these representatives are homogeneous of degree two on $\PT$ and $\PT^*$, respectively. The ambitwistor string vertex operators corresponding to these conformal primary gravitons are given by
\be\label{cgrvo+}
\begin{split}
\cV^{\veps}_{+,\Delta}(z,\bar{z})&=\int\frac{\d s}{s^3}\,\frac{\d t}{t^{3-\Delta}}\,\left(\left[\tilde{\lambda}\,\frac{\partial}{\partial\mu}\right]+\tilde{\rho}^{\dot\alpha}\,\rho^{\dot\beta}\,\frac{\partial^2}{\partial\mu^{\dot\alpha}\partial\mu^{\dot\beta}}\right)\bar\delta^2(z-s\,\lambda)\,\e^{\im\veps\,t\,s\,[\mu\,\bar z]} \\
 &=\int\frac{\d s}{s^2}\,\frac{\d t}{t^{2-\Delta}}\,\veps\,\left(\im\,[\tilde{\lambda}\,\bar{z}]-\veps\,s\,t\,[\tilde{\rho}\,\bar{z}]\,[\rho\,\bar{z}] \right)\bar\delta^2(z-s\,\lambda)\,\e^{\im\veps\,t\,s\,[\mu\,\bar z]}\,,
\end{split}
\ee
and
\be\label{cgrvo-}
\begin{split}
\cV^{\veps}_{-,\Delta}(z,\bar{z})&=\int\frac{\d \tilde{s}}{\tilde{s}^3}\,\frac{\d t}{t^{3-\Delta}}\,\left(\left\la\lambda\,\frac{\partial}{\partial\tilde\mu}\right\ra+\rho^{\alpha}\,\tilde{\rho}^{\beta}\,\frac{\partial^2}{\partial\tilde{\mu}^{\alpha}\partial\tilde{\mu}^{\beta}}\right)\bar{\delta}^{2}(\bar{z}-\tilde{s}\,\tilde{\lambda})\,\e^{\im\veps\,t\,\tilde{s}\,\la\tilde{\mu}\,z\ra} \\
 &=\int\frac{\d \tilde{s}}{\tilde{s}^2}\,\frac{\d t}{t^{2-\Delta}}\,\veps\,\left(\im\,\la\lambda\,z\ra-\veps\,\tilde{s}\,t\,\la\rho\,z\ra\,\la\tilde{\rho}\,z\ra \right)\bar\delta^2(\bar{z}-\tilde{s}\,\tilde{\lambda})\,\e^{\im\veps\,t\,\tilde{s}\,\la\tilde{\mu}\,z\ra}\,,
\end{split}
\ee
where the dependence of worldsheet variables on their location on $\Sigma$ has been suppressed. The families of vertex operators \eqref{cglvo+}, \eqref{cglvo-}, \eqref{cgrvo+}, \eqref{cgrvo-} and the simple worldsheet OPEs of the ambitwistor string are the basic tools for the remainder of this paper.


\section{Conformal primary basis: same helicity OPEs}\label{same_h}

Although it is well-known that correlation functions of the ambitwistor string produce field theory scattering amplitudes, there is \emph{a priori} no reason to expect that the worldsheet OPE between vertex operators will lead to anything meaningful on its own (i.e., outside of a correlation function), let alone have any relationship to the OPE limit in CCFT. This is because the scattering equations, which tie collinear limits with factorization, only emerge \emph{after} the worldsheet path integral for a complete correlation function is performed. Remarkably, in the next two sections we show that the \emph{worldsheet} OPEs between vertex operators of in the ambitwistor string directly reproduces the corresponding \emph{celestial} OPEs of gluons or gravitons in a CCFT. 

In this section we begin by considering gluon and graviton vertex operators of the same helicity. This is because the structure of the worldsheet OPEs is particularly simple for same-helicity vertex operators, providing an easy warm-up for the more subtle mixed-helicity calculations in Section~\ref{Sec:Mixed}.  


\subsection{Gluon-gluon OPE}\label{GluonOPE}

Without loss of generality\footnote{The OPE between negative helicity fields is given by the obvious `complex conjugate' of the result for positive helicity fields.}, consider the vertex operators for two positive helicity gluons in the conformal primary basis, $\cU^{\msf{a},\veps_i}_{+,\Delta_i}(z_i,\bar{z}_i)$ and $\cU^{\msf{b},\veps_j}_{+,\Delta_j}(z_j,\bar{z}_j)$ given by \eqref{cglvo+}. Since the wavefunctions in both vertex operators live on $\PT$, the only non-trivial contribution to the worldsheet OPE between them comes from the worldsheet current algebra via \eqref{wcaOPE}. The OPE is thus easily seen to give
\begin{multline}\label{shgl1}
\cU^{\msf{a},\veps_i}_{+,\Delta_i}\,\cU^{\msf{b},\veps_j}_{+,\Delta_j} \sim \int\d\sigma_i\,\frac{f^{\msf{abc}}\,j^{\msf{c}}(\sigma_j)}{\sigma_{ij}}\,\frac{\d s_i}{s_i}\,\frac{\d s_j}{s_j}\,\frac{\d t_i}{t_i^{2-\Delta_i}}\,\frac{\d t_j}{t_j^{2-\Delta_j}}\,\bar\delta^2\left(z_i-s_i\,\lambda(\sigma_i)\right) \\
\bar\delta^2\left(z_j-s_j\,\lambda(\sigma_j)\right)\,\exp\Big(\im\veps_i\,t_i\,s_i\,[\mu(\sigma_i)\,\bar z_i]+\im\veps_j\,t_j\,s_j\,[\mu(\sigma_j)\,\bar z_j]\Big)\,,
\end{multline}
where $\sigma_{ij}:=\sigma_i-\sigma_j$.

The first thing to observe about this expression is that in the limit where the insertion points on the worldsheet collide, $\sigma_{ij}\to0$, the holomorphic delta functions enforce $\la z_{i}\,z_{j}\ra\to 0$ or equivalently $z_{ij}\to0$. This tells us that the OPE limit on the worldsheet actually coincides with the holomorphic OPE limit (i.e., $z_{ij}\to0$) on the celestial sphere. While this is already a somewhat surprising fact, the right-hand-side of \eqref{shgl1} can be substantially simplified.

To begin, the scale integral over $s_i$ can be performed against one of the first holomorphic delta functions by projecting it onto an arbitrary basis of un-dotted spinors. This sets $s_{i}=\la\xi\,z_i\ra/\la\xi\,\lambda(\sigma_i)\ra$ for $\xi_{\alpha}$ not proportional to $z_{i\,\alpha}$ but otherwise arbitrary. This means that we are free to set $\xi_{\alpha}=\iota_{\alpha}=(0,1)$, for which $\la\iota\,z_i\ra=1$, leaving
\begin{multline}\label{shgl2}
 \cU^{\msf{a},\veps_i}_{+,\Delta_i}\,\cU^{\msf{b},\veps_j}_{+,\Delta_j} \sim-\int\d\sigma_i\,\frac{f^{\msf{abc}}\,j^{\msf{c}}(\sigma_j)}{\sigma_{ij}}\,\frac{\d s_j}{s_j}\,\frac{\d t_i}{t_i^{2-\Delta_i}}\,\frac{\d t_j}{t_j^{2-\Delta_j}}\,\la\iota\,\lambda(\sigma_i)\ra\,\bar{\delta}\!\left(\la z_{i}\,\lambda(\sigma_i)\ra\right)  \\
\bar\delta^2\left(z_j-s_j\,\lambda(\sigma_j)\right)\,\exp\left(\im\veps_i\,t_i\,\frac{[\mu(\sigma_i)\,\bar z_i]}{\la\iota\,\lambda(\sigma_i)\ra}+\im\veps_j\,t_j\,s_j\,[\mu(\sigma_j)\,\bar z_j]\right)\,,
\end{multline}
after performing the $s_i$ integral. Now, using the definition of the holomorphic delta \eqref{hdelta1}, we can integrate-by-parts with respect to the $\dbar$-operator acting on $\sigma_i$. This trades the holomorphic delta function in $\la z_{i}\,\lambda(\sigma_i)\ra$ to one in $\sigma_{ij}$:
\begin{multline}\label{shgl3}
 \cU^{\msf{a},\veps_i}_{+,\Delta_i}\,\cU^{\msf{b},\veps_j}_{+,\Delta_j} \sim f^{\msf{abc}}\,\int\d\sigma_i\,j^{\msf{c}}(\sigma_j)\,\bar{\delta}(\sigma_{ij})\,\frac{\d s_j}{s_j}\,\frac{\d t_i}{t_i^{2-\Delta_i}}\,\frac{\d t_j}{t_j^{2-\Delta_j}}\,\frac{\la\iota\,\lambda(\sigma_i)\ra}{\la z_{i}\,\lambda(\sigma_i)\ra}  \\
\bar\delta^2\left(z_j-s_j\,\lambda(\sigma_j)\right)\,\exp\left(\im\veps_i\,t_i\,\frac{[\mu(\sigma_i)\,\bar z_i]}{\la\iota\,\lambda(\sigma_i)\ra}+\im\veps_j\,t_j\,s_j\,[\mu(\sigma_j)\,\bar z_j]\right)\,.
\end{multline}
This shows that the OPE of worldsheet vertex operators actually localizes on the $\sigma_{ij}\to0$ region of the moduli space, and thus singles out the holomorphic OPE limit on the celestial sphere.

The worldsheet integral in $\sigma_i$ can now be performed against $\bar{\delta}(\sigma_{ij})$ to give
\begin{multline}\label{shgl4}
 \cU^{\msf{a},\veps_i}_{+,\Delta_i}\,\cU^{\msf{b},\veps_j}_{+,\Delta_j} \sim f^{\msf{abc}}\,\int j^{\msf{c}}(\sigma_j)\,\frac{\d s_j}{s_j}\,\frac{\d t_i}{t_i^{2-\Delta_i}}\,\frac{\d t_j}{t_j^{2-\Delta_j}}\,\frac{\la\iota\,\lambda(\sigma_j)\ra}{\la z_{i}\,\lambda(\sigma_j)\ra}  \\
\bar\delta^2\left(z_j-s_j\,\lambda(\sigma_j)\right)\,\exp\left(\im\veps_i\,t_i\,\frac{[\mu(\sigma_j)\,\bar z_i]}{\la\iota\,\lambda(\sigma_j)\ra}+\im\veps_j\,t_j\,s_j\,[\mu(\sigma_j)\,\bar z_j]\right)\,.
\end{multline}
This is further simplified by noting that on the support of the remaining holomorphic delta functions
\be\label{delsupport}
\frac{\la\iota\,\lambda(\sigma_j)\ra}{\la z_{i}\,\lambda(\sigma_j)\ra}=\frac{\la\iota\,z_j\ra}{\la z_{i}\,z_j\ra}=\frac{1}{z_{ij}}\,, \qquad \frac{1}{\la\iota\,\lambda(\sigma_j)\ra}=\frac{s_j}{\la\iota\,z_{j}\ra}=s_j\,.
\ee
Combining this with the Mellin parameter rescaling $t_{i}\mapsto t_{i}t_{j}$ leaves
\begin{multline}\label{shgl5}
 \cU^{\msf{a},\veps_i}_{+,\Delta_i}\,\cU^{\msf{b},\veps_j}_{+,\Delta_j} \sim \frac{f^{\msf{abc}}}{z_{ij}}\,\int j^{\msf{c}}(\sigma_j)\,\frac{\d s_j}{s_j}\,\frac{\d t_i}{t_i^{2-\Delta_i}}\,\frac{\d t_j}{t_j^{3-\Delta_i-\Delta_j}}\,\bar\delta^2\left(z_j-s_j\,\lambda(\sigma_j)\right) \\
 \exp\Big[\im\,t_j\,s_j\left(\veps_i\,t_i\,[\mu(\sigma_j)\,\bar z_i]+\veps_j\,[\mu(\sigma_j)\,\bar z_j]\right)\Big]\,.
\end{multline}
A bit of straightforward algebra reveals that the argument of the exponential appearing on the second line of this expression can be written as
\begin{multline}\label{exparrange}
\im\,t_j\,s_j\left(\veps_i\,t_i\,[\mu(\sigma_j)\,\bar z_i]+\veps_j\,[\mu(\sigma_j)\,\bar z_j]\right) \\ =\im\,s_{j}\,t_j\left(1+\frac{\veps_i}{\veps_j}\,t_i\right)\left(\frac{\veps_i\,t_i}{1+\frac{\veps_i}{\veps_j}\,t_i}[\mu(\sigma_j)\,\bar{z}_{ij}]+\veps_j\,[\mu(\sigma_j)\,\bar{z}_j]\right)\,,
\end{multline}
where $\bar z_{ij\,\dal} \equiv \bar z_{i\,\dal}-\bar z_{j\,\dal} = (0,\bar z_{ij})$. Accompanied by a rescaling $t_j\mapsto t_j |1+\frac{\veps_i}{\veps_j}t_i|^{-1}$, the worldsheet OPE becomes
\begin{multline}\label{shgl6}
 \cU^{\msf{a},\veps_i}_{+,\Delta_i}\,\cU^{\msf{b},\veps_j}_{+,\Delta_j} \sim \frac{f^{\msf{abc}}}{z_{ij}}\,\int j^{\msf{c}}(\sigma_j)\,\frac{\d s_j}{s_j}\,\frac{\d t_i\,t_i^{\Delta_i-2}}{|1+\frac{\veps_i}{\veps_j}\,t_i|^{\Delta_i+\Delta_j-2}}\,\frac{\d t_j}{t_j^{3-\Delta_i-\Delta_j}}\,\bar\delta^2\left(z_j-s_j\,\lambda(\sigma_j)\right) \\
 \exp\left[\im\,t_j\,s_j\left(\frac{\veps_i\,t_i}{|1+\frac{\veps_i}{\veps_j}\,t_i|}\,[\mu(\sigma_j)\,\bar{z}_{ij}]+\mathrm{sgn}(\veps_j+\veps_i\,t_i)\,[\mu(\sigma_j)\,\bar z_j]\right)\right]\,,
\end{multline}
where $\sgn(\cdots)$ is the sign function and the remaining integrals are over the worldsheet location parametrized by $\sigma_j$, the complex scaling parameter $s_j$, and two Mellin parameters $t_i,t_j\in\R_+$.

Now, it is straightforward to expand the first term in the exponential, leading to a final expression for the OPE:
\be\label{shgl7}
 \cU^{\msf{a},\veps_i}_{+,\Delta_i}\,\cU^{\msf{b},\veps_j}_{+,\Delta_j} \sim \frac{f^{\msf{abc}}}{z_{ij}}\,\sum_{m=0}^{\infty}\frac{1}{m!}\int_{0}^{\infty}\frac{\d t_i\,t_i^{\Delta_i+m-2}}{|1+\frac{\veps_i}{\veps_j}\,t_i|^{\Delta_i+\Delta_j+m-2}}\,\left(\frac{\veps_i\,\bar{z}_{ij}}{\veps_j}\right)^{m}\,\dbar_j^m\cU^{\msf{c},\sgn(\veps_j+\veps_i t_i)}_{+,\Delta_i+\Delta_j-1}(z_j,\bar{z}_j)\,,
\ee
where another positive helicity gluon vertex operator has been identified on the right-hand-side using the definition \eqref{cglvo+}. This is an \emph{exact} result for the worldsheet OPE: no singular terms have been neglected, and the right-hand-side of \eqref{shgl7} is literally equal to the right-hand-side of \eqref{shgl1}. The central claim is that this worldsheet OPE is equal to the celestial OPE between positive helicity gluons in any incoming/outgoing configuration.

\medskip

To see this, first consider the case where both gluons are incoming or outgoing: $\veps_i=\veps_j=\veps$. With this assumption, it is easy to see that \eqref{shgl7} gives 
\be\label{shglout}
\cU^{\msf{a},\veps}_{+,\Delta_i}\,\cU^{\msf{b},\veps}_{+,\Delta_j} \sim \frac{f^{\msf{abc}}}{z_{ij}}\,\sum_{m=0}^{\infty}B(\Delta_i+m-1,\Delta_j-1)\,\frac{\bar{z}_{ij}^m}{m!}\,\dbar_j^m\cU^{\msf{c},\veps}_{+,\Delta_i+\Delta_j-1}(z_j,\bar{z}_j)\,,
\ee
using one of the standard integral representations of the Beta function. As desired, this matches the OPE \eqref{shgluon} between gluons of the same helicity and orientation in CCFT in pure Yang-Mills theory. 

For the mixed incoming/outgoing configuration ($\veps_i=-\veps_j=\veps$), one must break the residual $\R_+$ integral in \eqref{shgl7} into two regions:
\begin{multline}\label{shglinout1}
\frac{f^{\msf{abc}}}{z_{ij}}\,\sum_{m=0}^{\infty}\frac{1}{m!}\int_{0}^{1}\frac{\d t_i\,t_i^{\Delta_i+m-2}}{(1-t_i)^{\Delta_i+\Delta_j+m-2}}\,\left(-\bar{z}_{ij}\right)^{m}\,\dbar_j^m\cU^{\msf{c},-\veps}_{+,\Delta_i+\Delta_j-1}(z_j,\bar{z}_j) \\
+\frac{f^{\msf{abc}}}{z_{ij}}\,\sum_{m=0}^{\infty}\frac{(-1)^{\Delta_i+\Delta_j}}{m!}\int_{1}^{\infty}\frac{\d t_i\,t_i^{\Delta_i+m-2}}{(1-t_i)^{\Delta_i+\Delta_j+m-2}}\,\left(\bar{z}_{ij}\right)^{m}\,\dbar_j^m\cU^{\msf{c},\veps}_{+,\Delta_i+\Delta_j-1}(z_j,\bar{z}_j)\,.
\end{multline}
The contribution from the first line is readily evaluated in terms of Beta functions
\be\label{shglinout2}
\frac{f^{\msf{abc}}}{z_{ij}}\,\sum_{m=0}^{\infty}B(\Delta_i+m-1,3-\Delta_i-\Delta_j-m)\frac{(-\bar{z}_{ij})^m}{m!}\,\dbar_j^m\cU^{\msf{c},-\veps}_{+,\Delta_i+\Delta_j-1}(z_j,\bar{z}_j)\,,
\ee
but the second line requires reparametrization by $t_{i}\mapsto t_{i}^{-1}$; this gives
\begin{multline}
\frac{f^{\msf{abc}}}{z_{ij}}\,\sum_{m=0}^{\infty}\frac{(-1)^{\Delta_i+\Delta_j}}{m!}\int_{0}^{1}\frac{\d t_i\,t_i^{\Delta_j-2}}{(1-t_i)^{\Delta_i+\Delta_j+m-2}}\,\bar{z}_{ij}^{m}\,\dbar_j^m\cU^{\msf{c},\veps}_{+,\Delta_i+\Delta_j-1}(z_j,\bar{z}_j) \\
=-\frac{f^{\msf{abc}}}{z_{ij}}\,\sum_{m=0}^{\infty}B(\Delta_j-1,3-\Delta_i-\Delta_j-m)\,\frac{\bar{z}_{ij}^m}{m!}\,\dbar_j^m\cU^{\msf{c},\veps}_{+,\Delta_i+\Delta_j-1}(z_j,\bar{z}_j)\,,
\end{multline}
where the sign $(-1)^{\Delta_i+\Delta_j}$ combines with the $\veps$-orientation of the gluon vertex operator to give an overall minus sign.

Putting all of this together, we have
\begin{multline}\label{shglinout3}
\cU^{\msf{a},\veps}_{+,\Delta_i}\,\cU^{\msf{b},-\veps}_{+,\Delta_j} \sim-\frac{f^{\msf{abc}}}{z_{ij}}\,\sum_{m=0}^{\infty}\frac{\bar{z}_{ij}^m}{m!}\,\dbar_j^m\left[B(\Delta_j-1,3-\Delta_i-\Delta_j-m)\,\cU^{\msf{c},\veps}_{+,\Delta_i+\Delta_j-1}(z_j,\bar{z}_j)\right. \\
\left.-(-1)^m\,B(\Delta_i+m-1,3-\Delta_i-\Delta_j-m)\,\cU^{\msf{c},-\veps}_{+,\Delta_i+\Delta_j-1}(z_j,\bar{z}_j) \right]\,,
\end{multline}
whose $m=0$ terms match the primary contributions to the incoming/outgoing celestial OPE for gluons in the literature~\cite{Fan:2019emx,Pate:2019lpp}. The $m>0$ terms here account for the infinite towers of $\SL(2,\R)$ descendant contributions to the OPE.


\subsection{Graviton-graviton OPE}\label{gravtion_same}

Next, consider the analogous calculation between two positive helicity gravitons in the conformal primary basis, represented by $\cV^{\veps_i}_{+,\Delta_i}(z_i,\bar{z}_i)$ and $\cV^{\veps_j}_{+,\Delta_j}(z_j,\bar{z}_j)$ given by \eqref{cgrvo+}. While this worldsheet OPE is more involved than a single contraction between worldsheet currents, it is still fairly straightforward to compute using the OPEs of fields \eqref{symp_bosons} and \eqref{wsferm}. In particular, there are no contractions involving the arguments of the holomorphic delta functions in either vertex operator. A somewhat tedious calculation gives
\begin{multline}\label{shgr1}
\cV^{\veps_i}_{+,\Delta_i}\,\cV^{\veps_j}_{+,\Delta_j}\sim \veps_{i}\,\veps_{j}\int  \frac{\d s_i}{s_i^3}\,\frac{\d s_j}{s_j^3}\,\frac{\d t_i}{t_{i}^{2-\Delta_i}}\,\frac{\d t_j}{t_j^{2-\Delta_j}}\,\frac{\bar{z}_{ij}}{\sigma_{ij}} \Big[-\im\,\veps_i\,t_i\,s_i\,[\tilde{\lambda}(\sigma_i)\,\bar{z}_i]-\im\,\veps_j\,t_j\,s_j\,[\tilde{\lambda}(\sigma_j)\,\bar{z}_j] \\
+\veps_i^2\,t_i^2\,s_i^2\,[\tilde{\rho}(\sigma_i)\,\bar{z}_{i}]\,[\rho(\sigma_i)\,\bar{z}_i] + \veps_j^2\,t_j^2\,s_j^2\,[\tilde{\rho}(\sigma_j)\,\bar{z}_{j}]\,[\rho(\sigma_j)\,\bar{z}_j] \\
+\veps_i\,\veps_j\,t_i\,t_j\,s_i\,s_j \left([\tilde{\rho}(\sigma_i)\,\bar{z}_i]\,[\rho(\sigma_j)\,\bar{z}_j]+[\tilde{\rho}(\sigma_j)\,\bar{z}_j]\,[\rho(\sigma_i)\,\bar{z}_i]\right)\Big]\,\d\sigma_i\,\d\sigma_j \\
\times\bar{\delta}^{2}(z_i-s_{i}\,\lambda(\sigma_i))\, \bar{\delta}^{2}(z_j-s_{j}\,\lambda(\sigma_j))\,\exp\Big(\im\veps_i\,t_i\,s_i\,[\mu(\sigma_i)\,\bar z_i]+\im\veps_j\,t_j\,s_j\,[\mu(\sigma_j)\,\bar z_j]\Big)\,.
\end{multline}
Note that while there could potentially have been double poles in this OPE, they cancel among each other thanks to the specific form of the graviton vertex operator. Once again, the holomorphic delta functions ensure that the $\sigma_{ij}\to0$ limit on the worldsheet corresponds to the $z_{ij}\to0$ limit on the celestial sphere.

At this point one follows the same strategy as in the gluon-gluon OPE: perform the $s_i$ integral against one of the holomorphic delta functions; integrate by parts in $\sigma_i$ to obtain $\bar{\delta}(\sigma_{ij})$; perform the $\sigma_i$ integral against this holomorphic delta function; and re-scale $t_i\mapsto t_i t_j$, $t_j\mapsto t_j |1+\frac{\veps_i}{\veps_j}t_i|^{-1}$. After taking these steps, \eqref{shgr1} is reduced to
\be\label{shgr2}
\begin{split}
\cV^{\veps_i}_{+,\Delta_i}\,\cV^{\veps_j}_{+,\Delta_j}&\sim \veps_{i}\veps_{j}\,\frac{\bar{z}_{ij}}{z_{ij}}\,\int\frac{\d s_j}{s_j^3}\,\frac{\d t_i\,t_i^{\Delta_i-2}}{|1+\frac{\veps_i}{\veps_j}\,t_i|^{\Delta_i+\Delta_j-2}}\,\frac{\d t_j}{t_j^{3-\Delta_i-\Delta_j}} \\
&\times\bar\delta^2\left(z_j-s_j\,\lambda(\sigma_j)\right) \left(\left[\tilde{\lambda}(\sigma_j)\,\frac{\partial}{\partial\mu(\sigma_j)}\right]+\tilde{\rho}^{\dot\alpha}\,\rho^{\dot\beta}\,\frac{\partial^2}{\partial\mu^{\dot\alpha}(\sigma_j)\partial\mu^{\dot\beta}(\sigma_j)}\right) \\
&\times\exp\left[\im\,t_j\,s_j\left(\frac{\veps_i\,t_i}{|1+\frac{\veps_i}{\veps_j}\,t_i|}\,[\mu(\sigma_j)\,\bar{z}_{ij}]+\sgn(\veps_j+\veps_i t_i)\,[\mu(\sigma_j)\,\bar z_j]\right)\right]\,.
\end{split}
\ee
It should be emphasized that after computing the initial worldsheet OPE, the details of this calculation do not differ from that of the gluon-gluon OPE in any significant way.

It is now straightforward to expand the first term in the exponential and identify a new positive helicity graviton vertex operator using \eqref{cgrvo+}. This results in a final expression for the same helicity graviton-graviton OPE:
\be\label{shgr3}
\cV^{\veps_i}_{+,\Delta_i}\,\cV^{\veps_j}_{+,\Delta_j}\sim \veps_{i}\veps_{j}\,\frac{\bar{z}_{ij}}{z_{ij}}\,\sum_{m=0}^{\infty}\frac{1}{m!}\int_{0}^{\infty}\frac{\d t_{i}\,t_i^{\Delta_i+m-2}}{|1+\frac{\veps_i}{\veps_j}\,t_i|^{\Delta_i+\Delta_j+m-2}}\left(\frac{\veps_i\,\bar{z}_{ij}}{\veps_j}\right)^m\,\dbar^{m}_{j}\cV^{\sgn(\veps_j+\veps_i t_i)}_{+,\Delta_i+\Delta_j}(z_j,\bar{z}_j)\,.
\ee
By taking different combinations of incoming/outgoing states, this relation captures the celestial OPE between positive helicity gravitons. The exact results are easily obtained using the same methods as before to give
\be\label{shgrout}
\cV^{\veps}_{+,\Delta_i}\,\cV^{\veps}_{+,\Delta_j}\sim \frac{\bar{z}_{ij}}{z_{ij}}\,\sum_{m=0}^{\infty}B(\Delta_i+m-1,\Delta_j-1)\,\frac{\bar{z}_{ij}^m}{m!}\,\dbar^{m}_{j}\cV^{\veps}_{+,\Delta_i+\Delta_j}(z_j,\bar{z}_j)\,,
\ee
and
\begin{multline}\label{shgrinout}
\cV^{\veps}_{+,\Delta_i}\,\cV^{-\veps}_{+,\Delta_j}\sim -\frac{\bar{z}_{ij}}{z_{ij}}\,\sum_{m=0}^{\infty}\frac{\bar{z}_{ij}^{m}}{m!}\,\dbar^{m}_{j}\left[B(\Delta_j-1,3-\Delta_i-\Delta_j-m)\,\cV^{\veps}_{+,\Delta_i+\Delta_j}(z_j,\bar{z}_j) \right. \\
\left. +(-1)^m\,B(\Delta_i+m-1,3-\Delta_i-\Delta_j-m)\,\cV^{-\veps}_{+,\Delta_i+\Delta_j}(z_j,\bar{z}_j)\right]\,,
\end{multline}
for both configurations. Sure enough, \eqref{shgrout} matches the full $\SL(2,\R)$ descendant CCFT OPE for same helicity, same orientation gravitons~\cite{Guevara:2021abz,Himwich:2021dau,Jiang:2021ovh}. The $m=0$ terms of \eqref{shgrinout} match the graviton primary contribution CCFT OPEs in the literature for same helicity, mixed orientation gravitons~\cite{Pate:2019lpp}; for $m>0$ we believe that this is the first calculation of the $\SL(2,\R)$ descendant contributions to the celestial OPE in this configuration. The veracity of the result is most easily established by Mellin transforming subleading collinear limits in momentum space; we illustrate how this works in Section~\ref{Sec:MomEig}.


\subsection{Gluon-graviton OPE}

Finally, consider the worldsheet OPE between a positive helicity graviton and a positive helicity gluon in the conformal primary basis, represented by $\cV^{\veps_i}_{+,\Delta_i}(z_i,\bar{z}_i)$ and $\cU^{\msf{a},\veps_j}_{+,\Delta_j}(z_j,\bar{z}_j)$, respectively. Once again, the structure of the worldsheet OPE is simple: the only singular contribution is a simple pole arising from the contraction of the $\tilde{\lambda}(\sigma_j)$ term in the graviton vertex operator with the exponential of the gluon vertex operator. The result is
\begin{multline}\label{shglgr1}
\cV^{\veps_i}_{+,\Delta_i}\,\cU^{\msf{a},\veps_j}_{+,\Delta_j}\sim \veps_i\veps_j\,\bar{z}_{ij}\,\int\frac{\d s_i}{s^2_i}\,\d s_j\,\frac{\d t_i}{t_i^{2-\Delta_i}}\,\frac{\d t_j}{t_j^{1-\Delta_j}}\,\frac{j^{\msf{a}}(\sigma_j)}{\sigma_{ij}}\,\d\sigma_i \\
\bar\delta^2\left(z_i-s_i\,\lambda(\sigma_i)\right) \,\bar\delta^2\left(z_j-s_j\,\lambda(\sigma_j)\right)\,\exp\Big(\im\veps_i\,t_i\,s_i\,[\mu(\sigma_i)\,\bar z_i]+\im\veps_j\,t_j\,s_j\,[\mu(\sigma_j)\,\bar z_j]\Big)\,,
\end{multline}
which one can proceed to further evaluate using the same steps as in the previous two cases. Proceeding in this way, one arrives at a final expression for the worldsheet OPE
\be\label{shglgr2}
\cV^{\veps_i}_{+,\Delta_i}\,\cU^{\msf{a},\veps_j}_{+,\Delta_j}\sim \veps_i\veps_j\,\frac{\bar{z}_{ij}}{z_{ij}}\sum_{m=0}^{\infty}\frac{1}{m!}\int_{0}^{\infty}\frac{\d t_{i}\,t_{i}^{\Delta_i+m-2}}{|1+\frac{\veps_i}{\veps_j}\,t_i|^{\Delta_i+\Delta_j+m-1}}\left(\frac{\veps_i\,\bar{z}_{ij}}{\veps_j}\right)^m\,\dbar^{m}_{j}\cU^{\msf{a},\sgn(\veps_j+\veps_i t_i)}_{+,\Delta_i+\Delta_j}(z_j,\bar{z}_j)\,,
\ee
capturing all incoming/outgoing configurations simultaneously.

As before, it is straightforward to extract the OPE results for particular configurations:
\be\label{shglgrout}
\cV^{\veps}_{+,\Delta_i}\,\cU^{\msf{a},\veps}_{+,\Delta_j}\sim\frac{\bar{z}_{ij}}{z_{ij}}\,\sum_{m=0}^{\infty}B(\Delta_i+m-1,\Delta_j)\,\frac{\bar{z}^m_{ij}}{m!}\,\dbar^m_{j}\cU^{\msf{a},\veps}_{+,\Delta_i+\Delta_j}(z_j,\bar{z}_j)\,,
\ee
and 
\begin{multline}\label{shglgrinout}
\cV^{\veps}_{+,\Delta_i}\,\cU^{\msf{a},-\veps}_{+,\Delta_j}\sim\frac{\bar{z}_{ij}}{z_{ij}}\,\dbar^m_j\left[B(\Delta_j,2-\Delta_i-\Delta_j-m)\,\cU^{\msf{a},\veps}_{+,\Delta_i+\Delta_j}(z_j,\bar{z}_j) \right. \\
\left. -(-1)^{m}\,B(\Delta_i+m-1,2-\Delta_i-\Delta_j-m)\,\cU^{\msf{a},-\veps}_{+,\Delta_i+\Delta_j}(z_j,\bar{z}_j)\right]\,.
\end{multline}
The first of these matches the $\SL(2,\R)$ descendant contribution to the celestial OPE of all incoming/outgoing gluons and gravitons in EYM~\cite{Guevara:2021abz,Himwich:2021dau,Jiang:2021ovh}, while the second expression generalizes the primary part of the OPE for mixed incoming/outgoing configurations~\cite{Pate:2019lpp}.


\section{Conformal primary basis: mixed helicity OPEs}
\label{Sec:Mixed}

Having demonstrated that the worldsheet and celestial OPEs coincide for gluons and gravitons of the same helicity, the next step is to investigate if this statement also holds true for mixed helicity configurations. On one hand, it seems obvious that it should: ambitwistor space is non-chiral, so why should the relation between worldsheet and celestial OPEs suddenly fail for mixed-helicity configurations? Yet even at the level of worldsheet correlation functions, the relationship between fields of different helicities is mysterious: in the scattering amplitude formulae produced by the 4d ambitwistor string, particles of different helicity only talk to each other through the (helicity-refined) scattering equations~\cite{Geyer:2014fka,Geyer:2016nsh}. These are fully on-shell objects which emerge only after performing the path integral for the worldsheet CFT.

The OPEs we consider are essentially off-shell -- since they are not part of a correlator -- so something could go wrong for mixed-helicity configurations. Sure enough, if one proceeds naively the calculation quickly breaks down. However, this is simply because the worldsheet OPE localises on a region of moduli space where the representation of the vertex operators is not valid. This is remedied by appropriately reparametrizing (i.e., going to a different coordinate patch on moduli space), after which the calculation proceeds analogously to the same helicity case, and agreement with the celestial OPE again emerges.


\subsection{Gluon-gluon OPE}
\label{sec:gluon+-}

As in the same-helicity case, the mixed-helicity gluon-gluon OPE is the simplest to deal with but also captures all essential features of the other particle configurations. Take one positive and one negative helicity gluon vertex operator in the ambitwistor string; these are represented by $\cU^{\msf{a},\veps_i}_{+,\Delta_i}(z_i,\bar{z}_i)$ from \eqref{cglvo+} and $\cU^{\msf{b},\veps_j}_{-,\Delta_j}(z_j,\bar{z}_j)$ from \eqref{cglvo-}, respectively. While there is still a simple pole contribution to the OPE from the worldsheet current algebra, the key difference from the same-helicity case is that there are now \emph{infinitely} many additional contractions between the wavefunction parts of the vertex operators.

In particular, we need to take into account the worldsheet OPEs between 
\be\label{mhwf1}
\bar{\delta}^{2}\!\left(z_i-s_i\,\lambda(\sigma_i)\right) \e^{\im\veps_i\,t_i\,s_i\,[\mu(\sigma_i)\,\bar{z}_i]}\,\bar{\delta}^{2}\!\left(\bar{z}_j-\tilde{s}_j\,\tilde{\lambda}(\sigma_j)\right) \e^{\im\veps_j\,t_j\,\tilde{s}_j\,\la\tilde{\mu}(\sigma_j)\,z_j\ra}\,.
\ee
The OPE between ambitwistor fields \eqref{symp_bosons} means that the exponential at $\sigma_i$ can contract into the holomorphic delta function at $\sigma_j$, and \emph{vice versa}; since the holomorphic delta functions \eqref{hdelta2} go like the reciprocal of their argument, there are infinitely many such contractions. Fortunately, these can be re-summed by representing the holomorphic delta functions as
\be\label{mhwf2}
\begin{split}
\bar{\delta}^{2}\!\left(z_i-s_i\,\lambda(\sigma_i)\right)&=\int_{\C^2}\frac{\d^{2}\msf{m}}{(2\pi)^2}\, \e^{\im\,\la\msf{m}\,z_i\ra-\im\,s_i\,\la\msf{m}\,\lambda(\sigma_i)\ra}\,, \\
\bar{\delta}^{2}\!\left(\bar{z}_j-\tilde{s}_j\,\tilde{\lambda}(\sigma_j)\right)&=\int_{\C^2}\frac{\d^{2}\tilde{\msf{m}}}{(2\pi)^2}\,\e^{\im\,[\tilde{\msf{m}}\,\bar{z}_j]-\im\,\tilde{s}_j\,[\tilde{\msf{m}}\,\tilde{\lambda}(\sigma_j)]}\,,
\end{split}
\ee
and applying well-known formulae for Wick contractions between exponential operators (e.g., \cite{Polchinski:1998rq}). For instance, 
\be\label{mhwf3}
\e^{\im\veps_i\,t_i\,s_i\,[\mu(\sigma_i)\,\bar{z}_i]}\,\e^{-\im\,\tilde{s}_j\,[\tilde{\msf{m}}\,\tilde{\lambda}(\sigma_j)]}\sim \exp\left(\im\,\frac{\veps_i\,t_i\,s_i\,\tilde{s}_j}{\sigma_{ij}}\,[\tilde{\msf{m}}\,\bar{z}_i]\right)\,:\e^{\im\veps_i\,t_i\,s_i\,[\mu(\sigma_i)\,\bar{z}_i]}\,\e^{-\im\,\tilde{s}_j\,[\tilde{\msf{m}}\,\tilde{\lambda}(\sigma_j)]}:\,,
\ee
where $:(\cdots):$ indicates normal-ordering.

After computing these exponentiated contributions, one arrives at the expression for the worldsheet OPE:
\begin{multline}\label{mhgl1}
\cU^{\msf{a},\veps_i}_{+,\Delta_i}\,\cU^{\msf{b},\veps_j}_{-,\Delta_j} \sim\int \d\sigma_i\,\frac{\d s_i}{s_i}\,\frac{\d\tilde{s}_j}{\tilde{s}_j}\,\frac{\d t_i}{t_i^{2-\Delta_i}}\,\frac{\d t_j}{t_j^{2-\Delta_j}}\,\frac{f^{\msf{abc}}\,j^{c}(\sigma_j)}{\sigma_{ij}}\,\e^{\im\veps_i\,t_i\,s_i\,[\mu(\sigma_i)\,\bar{z}_i]}\,\e^{\im\veps_j\,t_j\,\tilde{s}_j\,\la\tilde{\mu}(\sigma_j)\,z_j\ra}\\
\bar{\delta}^{2}\!\left(z_i-s_{i}\,\lambda(\sigma_i)-\frac{\veps_j\,t_j\,s_i\,\tilde{s}_j\,z_j}{\sigma_{ij}}\right)\,\bar{\delta}^{2}\!\left(\bar{z}_j-\tilde{s}_j\,\tilde{\lambda}(\sigma_j)+\frac{\veps_i\,t_i\,s_i\,\tilde{s}_j\,\bar{z}_i}{\sigma_{ij}}\right)\,,
\end{multline}
with all remaining worldsheet fields implicitly normal-ordered so that there are no further singular contributions. Clearly, the structure of the constraints imposed by the holomorphic delta functions is substantially changed from the same helicity case at the same stage \eqref{shgl1} of the calculation. In the limit $\sigma_{ij}\to0$, the arguments of the holomorphic delta functions become singular unless one of the scaling parameters $s_i$ or $\tilde{s}_j$ simultaneously vanishes.

Suppose that $s_i\to0$ as $\sigma_{ij}\to0$, with the ratio $s_i/\sigma_{ij}$ remaining finite; without loss of generality we can assume that $s_i/\sigma_{ij}\to1$. In this case, the first delta functions impose $z_{ij}\to0$, so the holomorphic celestial limit emerges. If instead we took $\tilde{s}_j\to0$ while $\tilde{s}_j/\sigma_{ij}\to1$, the second delta functions impose $\bar{z}_{ij}\to0$. So requiring that the arguments of the holomorphic delta functions remain finite in the $\sigma_{ij}\to0$ means that one automatically recovers both the holomorphic and anti-holomorphic OPE limits on the celestial sphere. In other words, the limit where insertion points collide on the worldsheet again coincides with the collision of insertion points in CCFT.

Unfortunately, we cannot simply follow the steps of the same-helicity calculation at this point. This would mean performing the scale integrals in $s_i$ or $\tilde{s}_j$ followed by an integration-by-parts to localise on $\sigma_{ij}\to0$. But we have just seen that this will require the vanishing of one of these scale parameters, which are affine (i.e., valued in $\C^*$) and \emph{not allowed to be zero}. In other words, the choice of scale parameters $s_i,\tilde{s}_j$ for the mixed-helicity OPE breaks down at the boundary divisor $\sigma_{ij}\to0$ of the moduli space. This means that a different parametrization is required to push the calculation further.

\medskip

Luckily, the required reparametrizations are remarkably simple and clearly suggested by the above observations. First, consider the reparametrization $s_i\mapsto s_i\,\sigma_{ij}$; on this new patch \eqref{mhgl1} becomes
\begin{multline}\label{mhgl2}
\int \d\sigma_i\,\frac{\d s_i}{s_i}\,\frac{\d\tilde{s}_j}{\tilde{s}_j}\,\frac{\d t_i}{t_i^{2-\Delta_i}}\,\frac{\d t_j}{t_j^{2-\Delta_j}}\,\frac{f^{\msf{abc}}\,j^{\msf{c}}(\sigma_j)}{\sigma_{ij}}\,\e^{\im\veps_i\,t_i\,s_i\,\sigma_{ij}\,[\mu(\sigma_i)\,\bar{z}_i]}\,\e^{\im\veps_j\,t_j\,\tilde{s}_j\,\la\tilde{\mu}(\sigma_j)\,z_j\ra}\\
\bar{\delta}^{2}\!\left(z_i-\sigma_{ij}\,s_{i}\,\lambda(\sigma_i)-\veps_{j}\,t_j\,s_i\,\tilde{s}_j\,z_j\right)\,\bar{\delta}^{2}\!\left(\bar{z}_j-\tilde{s}_j\,\tilde{\lambda}(\sigma_j)+\veps_i\,t_i\,s_i\,\tilde{s}_j\,\bar{z}_i\right)\,,
\end{multline}
and the singularity in the holomorphic delta functions as $\sigma_{ij}\to0$ is removed. Now we follow the blueprint of the same-helicity calculation by performing the $s_i$-integral against the first set of holomorphic delta functions. This gives
\begin{multline}\label{mhgl3}
 -\int \frac{\d\sigma_i}{s_i^{\star}}\,\frac{\d\tilde{s}_j}{\tilde{s}_j}\,\frac{\d t_i}{t_i^{2-\Delta_i}}\,\frac{\d t_j}{t_j^{2-\Delta_j}}\,\frac{f^{\msf{abc}}\,j^{\msf{c}}(\sigma_j)}{\sigma_{ij}}\,\e^{\im\veps_i\,t_i\,s^{\star}_i\,\sigma_{ij}\,[\mu(\sigma_i)\,\bar{z}_i]}\,\e^{\im\veps_j\,t_j\,\tilde{s}_j\,\la\tilde{\mu}(\sigma_j)\,z_j\ra} \\
 \bar{\delta}\!\left(\sigma_{ij}\,\la z_i\,\lambda(\sigma_i)\ra+\veps_j\,t_j\,\tilde{s}_j\,z_{ij}\right)\,\bar{\delta}^{2}\!\left(\bar{z}_j-\tilde{s}_j\,\tilde{\lambda}(\sigma_j)+\veps_i\,t_i\,s_i^{\star}\,\tilde{s}_j\,\bar{z}_i\right)\,,
\end{multline}
where
\be\label{sstar}
s_i^{\star}:=\frac{1}{\sigma_{ij}\,\la\iota\,\lambda(\sigma_i)\ra+\veps_j\,t_j\,\tilde{s}_j}\,,
\ee
is the value of $s_i$ which is fixed upon integration against the holomorphic delta function.

It is now possible to integrate-by-parts in $\sigma_i$, localising on the $\sigma_{ij}=0$ pole. After rescaling $t_i\mapsto t_i\,t_j$ this leaves
\be\label{mhgl4}
 \frac{f^{\msf{abc}}}{z_{ij}}\,\int\frac{\d\tilde{s}_j}{\tilde{s}_j}\,\frac{\d t_i}{t_i^{2-\Delta_i}}\,\frac{\d t_j}{t_j^{3-\Delta_i-\Delta_j}}\,j^{\msf{c}}(\sigma_j)\,
 \bar{\delta}^{2}\!\left(\bar{z}_j-\tilde{s}_j\,\tilde{\lambda}(\sigma_j)+\frac{\veps_i}{\veps_j}\,t_i\,\bar{z}_i\right)\,\e^{\im\veps_j\,t_j\,\tilde{s}_j\,\la\tilde{\mu}(\sigma_j)\,z_j\ra}\,.
\ee
Performing two further rescalings of the remaining integration parameters
\be\label{rescales}
\tilde{s}_j\mapsto \left(1+\frac{\veps_i}{\veps_j}\,t_i\right)\,\tilde{s}_j\,, \qquad t_j\mapsto \frac{t_j}{|1+\frac{\veps_i}{\veps_j}\,t_i|}\,,
\ee
enables one to recast the argument of the remaining holomorphic delta functions in a suggestive form:
\begin{multline}\label{mhgl5}
 \frac{f^{\msf{abc}}}{z_{ij}}\,\int\frac{\d\tilde{s}_j}{\tilde{s}_j}\,\frac{\d t_i\,t_i^{\Delta_i-2}}{|1+\frac{\veps_i}{\veps_j}\,t_i|^{\Delta_i+\Delta_j}}\,\frac{\d t_j}{t_j^{3-\Delta_i-\Delta_j}}\,j^{\msf{c}}(\sigma_j) \\
 \bar{\delta}^{2}\!\left(\bar{z}_j-\tilde{s}_j\,\tilde{\lambda}(\sigma_j)+\frac{\veps_i\,t_i}{\veps_j+\veps_i\,t_i}\,\bar{z}_{ij}\right)\,\e^{\im\,\sgn(\veps_j+\veps_i t_i)\,t_j\,\tilde{s}_j\,\la\tilde{\mu}(\sigma_j)\,z_j\ra}\,.
\end{multline}
Finally, the holomorphic delta function can be expanded (e.g., by exponentiating as in \eqref{mhwf2} and expanding the exponential in $\bar{z}_{ij}$) to give a final formula
\be\label{mhgl6}
\frac{f^{\msf{abc}}}{z_{ij}}\,\sum_{m=0}^{\infty}\frac{1}{m!}\,\int_{0}^{\infty}\frac{\d t_i\,t_i^{\Delta_i+m-2}}{|1+\frac{\veps_i}{\veps_j}\,t_i|^{\Delta_i+\Delta_j}} \left(\frac{\veps_i\,\bar{z}_{ij}}{\veps_j+\veps_i\,t_i}\right)^m\,\dbar^m_j\cU^{\msf{c},\sgn(\veps_j+\veps_i t_i)}_{-,\Delta_i+\Delta_j-1}(z_j,\bar{z_j})\,,
\ee
for the contribution to the OPE from this reparametrized coordinate patch.

\medskip

Of course, there is a second contribution to the worldsheet OPE coming from another coordinate patch on moduli space associated with the reparametrization $\tilde{s}_j\mapsto \tilde{s}_j\,\sigma_{ij}$. The computation of this contribution follows steps practically identical to those above, and summing the two together gives the full worldsheet OPE between conformal primary gluons of different helicity:
\begin{multline}\label{mhgl7}
 \cU^{\msf{a},\veps_i}_{+,\Delta_i}\,\cU^{\msf{b},\veps_j}_{-,\Delta_j} \sim\frac{f^{\msf{abc}}}{z_{ij}}\,\sum_{m=0}^{\infty}\frac{1}{m!}\,\int_{0}^{\infty}\frac{\d t_i\,t_i^{\Delta_i+m-2}}{|1+\frac{\veps_i}{\veps_j}\,t_i|^{\Delta_i+\Delta_j}} \left(\frac{\veps_i\,\bar{z}_{ij}}{\veps_j+\veps_i\,t_i}\right)^m\,\dbar^m_j\cU^{\msf{c},\sgn(\veps_j+\veps_i t_i)}_{-,\Delta_i+\Delta_j-1}(z_j,\bar{z_j}) \\
 +\frac{f^{\msf{abc}}}{\bar{z}_{ji}}\,\sum_{m=0}^{\infty}\frac{1}{m!}\,\int_{0}^{\infty}\frac{\d t_j\,t_j^{\Delta_j+m-2}}{|1+\frac{\veps_j}{\veps_i}\,t_j|^{\Delta_i+\Delta_j}} \left(\frac{\veps_j\,z_{ji}}{\veps_i+\veps_j\,t_j}\right)^m\,\partial^m_i\,\cU^{\msf{c},\sgn(\veps_i+\veps_j t_j)}_{+,\Delta_i+\Delta_j-1}(z_i,\bar{z}_i)\,.
\end{multline}
The mixed-helicity worldsheet OPE produces two classes of terms in the CCFT: those with a holomorphic singularity on the celestial sphere and those with an anti-holomorphic singularity. These couple to the two distinct helicities which can appear as new states in the celestial OPE.

As before, it is straightforward to evaluate \eqref{mhgl7} on specific incoming/outgoing configurations. For instance, 
\begin{multline}\label{mhglout}
 \cU^{\msf{a},\veps}_{+,\Delta_i}\,\cU^{\msf{b},\veps}_{-,\Delta_j} \sim\frac{f^{\msf{abc}}}{z_{ij}}\,\sum_{m=0}^{\infty}B(\Delta_i+m-1,\Delta_j+1)\,\frac{\bar{z}_{ij}^m}{m!}\,\dbar^m_j\cU^{\msf{c},\veps}_{-,\Delta_i+\Delta_j-1}(z_j,\bar{z_j}) \\
 +\frac{f^{\msf{abc}}}{\bar{z}_{ji}}\,\sum_{m=0}^{\infty} B(\Delta_j+m-1,\Delta_i+1)\,\frac{z_{ji}^m}{m!}\,\partial^m_i\,\cU^{\msf{c},\veps}_{+,\Delta_i+\Delta_j-1}(z_i,\bar{z}_i)\,,
\end{multline}
which matches the results in the literature~\cite{Pate:2019lpp} at the level of the primary contributions ($m=0$) and includes the full tower of $\SL(2,\R)$ descendants for each chirality in pure Yang-Mills theory. More generally, in the presence of gravity this relation is modified by terms corresponding to the two gluons fusing into a graviton~\cite{Pate:2019lpp}. Such terms are mediated in the ambitwistor string by the level of the worldsheet current algebra, which we have switched off precisely to decouple non-unitary gravitational degrees of freedom. As a result, these gluon-fusion terms are the only contributions to the celestial OPE which we are currently unable to produce with the ambitwistor string.


\subsection{Graviton-graviton OPE}

In contrast to the same-helicity case, the initial structure of the graviton-graviton worldsheet OPE is even simpler than that of the gluon. This is because the only non-trivial contractions are those between the exponentials and the holomorphic delta functions; there are no contractions between the pre-exponential parts of graviton vertex operators \eqref{cgrvo+} and \eqref{cglvo-} of opposite helicity. So the starting point of the computation is simply
\begin{multline}\label{mhgr1}
 \cV^{\veps_i}_{+,\Delta_i}\,\cV^{\veps_j}_{-,\Delta_j}\sim \veps_i\veps_j\,\int \frac{\d s_i}{s^2_i}\,\frac{\d\tilde{s}_j}{\tilde{s}^2_j}\,\frac{\d t_i}{t_i^{2-\Delta_i}}\,\frac{\d t_j}{t_j^{2-\Delta_j}}\,\e^{\im\veps_i\,t_i\,s_i\,[\mu(\sigma_i)\,\bar{z}_i]}\,\e^{\im\veps_j\,t_j\,\tilde{s}_j\,\la\tilde{\mu}(\sigma_j)\,z_j\ra}\\
 \left(\im\,[\tilde{\lambda}(\sigma_i)\,\bar{z}_i]-\veps_i\,s_i\,t_i\,[\tilde{\rho}(\sigma_i)\,\bar{z}_i]\,[\rho(\sigma_i)\,\bar{z}_i] \right)  \left(\im\,\la\lambda(\sigma_j)\,z_j\ra-\veps_j\,\tilde{s}_j\,t_j\,\la\rho(\sigma_j)\,z_j\ra\,\la\tilde{\rho}(\sigma_j)\,z_j\ra \right)\\
\bar{\delta}^{2}\!\left(z_i-s_{i}\,\lambda(\sigma_i)-\frac{\veps_j\,t_j\,s_i\,\tilde{s}_j\,z_j}{\sigma_{ij}}\right)\,\bar{\delta}^{2}\!\left(\bar{z}_j-\tilde{s}_j\,\tilde{\lambda}(\sigma_j)+\frac{\veps_i\,t_i\,s_i\,\tilde{s}_j\,\bar{z}_i}{\sigma_{ij}}\right)\,.
\end{multline}
To proceed, one follows exactly the same steps as in the gluon-gluon calculation: namely, evaluating on two reparametrized patches ($s_i\mapsto s_i\,\sigma_{ij}$ and $\tilde{s}_j\mapsto\tilde{s}_j\,\sigma_{ij}$) and performing the usual sequence of integrations and rescalings. 

The final result is
\begin{multline}\label{mhgr2}
 \cV^{\veps_i}_{+,\Delta_i}\,\cV^{\veps_j}_{-,\veps_j}\sim \veps_i\veps_j\,\frac{\bar{z}_{ij}}{z_{ij}}\,\sum_{m=0}^{\infty}\frac{1}{m!}\int_{0}^{\infty}\frac{\d t_i\,t_i^{\Delta_i+m-2}}{|1+\frac{\veps_i}{\veps_j}\,t_i|^{\Delta_i+\Delta_j+2}} \left(\frac{\veps_i\,\bar{z}_{ij}}{\veps_j+\veps_i\,t_i}\right)^m\,\dbar^m_j\cV^{\sgn(\veps_j+\veps_i t_i)}_{-,\Delta_i+\Delta_j}(z_j,\bar{z_j}) \\
 +\veps_i\veps_j\,\frac{z_{ji}}{\bar{z}_{ji}}\,\sum_{m=0}^{\infty}\frac{1}{m!}\,\int_{0}^{\infty}\frac{\d t_j\,t_j^{\Delta_j+m-2}}{|1+\frac{\veps_j}{\veps_i}\,t_j|^{\Delta_i+\Delta_j+2}} \left(\frac{\veps_j\,z_{ji}}{\veps_i+\veps_j\,t_j}\right)^{m}\,\partial^m_i\,\cV^{\sgn(\veps_i+\veps_j t_j)}_{+,\Delta_i+\Delta_j}(z_i,\bar{z}_i)\,.
\end{multline}
This can be evaluated more explicitly for specific incoming/outgoing configurations; for instance
\begin{multline}\label{mhgrout}
 \cV^{\veps}_{+,\Delta_i}\,\cV^{\veps}_{-,\Delta_j}\sim \frac{\bar{z}_{ij}}{z_{ij}}\,\sum_{m=0}^{\infty} B(\Delta_i+m-1,\Delta_j+3)\,\frac{\bar{z}_{ij}^m}{m!}\,\dbar^m_j\cV^{\veps}_{-,\Delta_i+\Delta_j}(z_j,\bar{z_j}) \\
 +\frac{z_{ji}}{\bar{z}_{ji}}\,\sum_{m=0}^{\infty} B(\Delta_j+m-1,\Delta_i+3)\,\frac{z_{ji}^m}{m!}\,\partial^m_i\,\cV^{\veps}_{+,\Delta_i+\Delta_j}(z_i,\bar{z}_i)\,.
\end{multline}
Once again, this matches the known result for the primary ($m=0$) contributions~\cite{Pate:2019lpp} while also including the infinite tower of $\SL(2,\R)$ descendants.


\subsection{Gluon-graviton OPE}

Finally, we consider the mixed-helicity OPE between graviton and gluon vertex operators, represented by $\cV^{\veps_i}_{+,\Delta_i}(z_i,\bar{z}_i)$ and $\cU^{\msf{a},\veps_j}_{-,\Delta_j}(z_j,\bar{z}_j)$, respectively. As in the graviton-graviton case, the only non-trivial part of the worldsheet OPE comes from contractions between exponentials and holomorphic delta functions, initially giving
\begin{multline}\label{mhglgr1}
\cV^{\veps_i}_{+,\Delta_i}\,\cU^{\msf{a},\veps_j}_{-,\Delta_j}\sim \veps_i\,\int \frac{\d s_i}{s^2_i}\,\frac{\d\tilde{s}_j}{\tilde{s}_j}\,\frac{\d t_i}{t_i^{2-\Delta_i}}\,\frac{\d t_j}{t_j^{2-\Delta_j}}\,\e^{\im\veps_i\,t_i\,s_i\,[\mu(\sigma_i)\,\bar{z}_i]}\,\e^{\im\veps_j\,t_j\,\tilde{s}_j\,\la\tilde{\mu}(\sigma_j)\,z_j\ra}\\
 \left(\im\,[\tilde{\lambda}(\sigma_i)\,\bar{z}_i]-\veps_i\,s_i\,t_i\,[\tilde{\rho}(\sigma_i)\,\bar{z}_i]\,[\rho(\sigma_i)\,\bar{z}_i] \right)  \, j^{\msf{a}}(\sigma_j)\\
\bar{\delta}^{2}\!\left(z_i-s_{i}\,\lambda(\sigma_i)-\frac{\veps_j\,t_j\,s_i\,\tilde{s}_j\,z_j}{\sigma_{ij}}\right)\,\bar{\delta}^{2}\!\left(\bar{z}_j-\tilde{s}_j\,\tilde{\lambda}(\sigma_j)+\frac{\veps_i\,t_i\,s_i\,\tilde{s}_j\,\bar{z}_i}{\sigma_{ij}}\right)\,.
\end{multline}
Following the same prescription as before, one discovers that the OPE is only non-trivial on the $s_{i}\mapsto s_i\,\sigma_{ij}$ patch; on the $\tilde{s}_j\mapsto\tilde{s}_j\,\sigma_{ij}$ patch the right-hand-side of \eqref{mhglgr1} is actually regular in $\sigma_{ij}$.

Thus, there is only one chirality of contribution to the OPE:
\be\label{mhglgr2}
\cV^{\veps_i}_{+,\Delta_i}\,\cU^{\msf{a},\veps_j}_{-,\Delta_j}\sim \veps_i\,\frac{\bar{z}_{ij}}{z_{ij}}\,\sum_{m=0}^{\infty}\frac{1}{m!}\int_{0}^{\infty}\frac{\d t_{i}\,t_i^{\Delta_i+m-2}}{|1+\frac{\veps_i}{\veps_j}\,t_i|^{\Delta_i+\Delta_j+1}} \left(\frac{\veps_i\,\bar{z}_{ij}}{\veps_j+\veps_i\,t_i}\right)^{m}\,\dbar^{m}_{j}\cU^{\msf{a},\sgn(\veps_j+\veps_i\,t_i)}_{-,\Delta_i+\Delta_j}(z_j,\bar{z}_j)\,,
\ee
which is consistent with helicity and colour-charge conservation. Of course, if one instead considered the $\cV^{\veps_i}_{-,\Delta_i}\,\cU^{\msf{a},\veps_j}_{+,\Delta_j}$ configuration then the resulting expression would be the `complex conjugate' of this. Once more, it is easy to see that this expression reproduces the primary contributions to the gluon-graviton celestial OPE in the literature~\cite{Pate:2019lpp} as well as generalizing them to include $\SL(2,\R)$ descendants.


\section{Holographic symmetry algebras}
\label{Sec:Sym}

One upshot of working with ambitwistor strings -- in contrast to twistor strings~\cite{Witten:2003nn,Berkovits:2004hg,Skinner:2013xp} or twistor sigma models~\cite{Adamo:2021bej,Adamo:2021lrv} -- is the ability to provide a parity-symmetric treatment of asymptotic symmetries associated to various conformally soft limits. In this section, we give concrete worldsheet vertex operator realizations of the Kac-Moody currents generating infinite-dimensional symmetries associated with the conformally soft gauge and gravity sectors. This also allows us to derive the action of soft gluons and gravitons on hard gluons and gravitons of any helicity in CCFT. 


\subsection{Soft gluon symmetries}

Soft gluon symmetries are associated to gauge transformations of holomorphic vector bundles on twistor space or dual twistor space, pulled back to ambitwistor space. Their generators can then be found from the conformally soft limits of gluon vertex operators. Here it will be easiest to work with a fully integrated form of the vertex operators, where a conformal primary positive helicity gluon has vertex operator
\be
\cU^{\msf{a},\,\veps}_{+,\Delta}(z,\bar z) = \int j^\msf{a}(\sigma)\,\bar\delta_{\Delta-1}\bigl(\la\lambda(\sigma)\,z\ra\bigr)\,\frac{(-\im\,\veps)^{1-\Delta}\,\Gamma(\Delta-1)}{[\mu(\sigma)\,\bar z]^{\Delta-1}}\,,
\ee 
where
\be
\bar\delta_{m}(\la\lambda\,\kappa\ra) := \frac{\la\iota\,\lambda\ra^{m+1}}{\la\iota\,\kappa\ra^{m+1}}\,\bar\delta(\la\lambda\,\kappa\ra) = \int_{\C^*}\frac{\d s}{s^{m+1}}\,\bar\delta^2(\kappa-s\lambda)
\ee
is a delta function of weight $m$ in $\lambda_\al$ and $-m-2$ in $\kappa_\al$. Taking $\veps=+1$ without loss of generality, the conformally soft gluons are defined as residues at the poles $\Delta = k \in \{1,0,-1,\dots\}$ in this vertex operator,
\be
\begin{split}
R^{k,\msf{a}}(z,\bar z) := \text{Res}_{\Delta=k}\,\cU^{\msf{a},+}_{+,\Delta}(z,\bar z) &= \int\frac{\im^{1-k}\,j^\msf{a}(\sigma)}{(1-k)!}\,[\mu(\sigma)\,\bar z]^{1-k}\,\bar\delta_{k-1}\bigl(\la\lambda(\sigma)\,z\ra\bigr)\\
&= \frac{1}{2\pi\im}\oint\frac{\im^{1-k}\,j^\msf{a}(\sigma)}{(1-k)!}\,\frac{[\mu(\sigma)\,\bar z]^{1-k}\,\la\iota\,\lambda(\sigma)\ra^{k}}{\la\lambda(\sigma)\,z\ra}\,.
\end{split}
\ee
Let us employ the index relabeling $k=3-2p$ and expand $[\mu\,\bar z] = \mu^{\dot0}+\bar z\,\mu^{\dot1}$. Then a binomial expansion in $\bar z$ gives
\be
R^{3-2p,\msf{a}}(z,\bar z) = \sum_{m=1-p}^{p-1}\frac{\bar z^{p-m-1}\,S^{p,\msf{a}}_m(z)}{\Gamma(p-m)\,\Gamma(p+m)}\,,
\ee
with currents $S^{p,\msf{a}}_m(z)$ defined by
\begin{align}
S^{p,\msf{a}}_m(z) &=  \frac{\im^{2p-2}}{2\pi\im}\oint\frac{j^\msf{a}(\sigma)\,g^p_m(\sigma)}{\la\iota\,\lambda(\sigma)\ra^{2p-3}\,\la\lambda(\sigma)\,z\ra}\,,\label{Spm}\\
g^p_m &:= (\mu^{\dot0})^{p+m-1}\,(\mu^{\dot1})^{p-m-1}\,.\label{gpm}
\end{align}
These currents generate gauge transformations localized at the $\lambda_\al = z_\al$ fiber of twistor space (any such fiber is diffeomorphic to a copy of space-time). 

By virtue of the current algebra OPE and the manipulations discussed in the previous sections, their algebra is easily found to be
\be
S^{p,\msf{a}}_m(z)\,S^{q,\msf{b}}_n(z') \sim \frac{\im^{2(p+q-1)-2}}{2\pi\im\,(z-z')}\oint\frac{f^\msf{abc}\,j^\msf{c}(\sigma)\,g^p_m(\sigma)\,g^q_n(\sigma)}{\la\iota\,\lambda(\sigma)\ra^{2(p+q-1)-3}\,\la\lambda(\sigma)\,z'\ra}\,.
\ee
Using $g^p_m\,g^q_n = g^{p+q-1}_{m+n}$, this is recognized to be the celestial OPE
\be
S^{p,\msf{a}}_m(z)\,S^{q,\msf{b}}_n(z') \sim \frac{f^\msf{abc}}{z-z'}\;S^{p+q-1,\msf{c}}_{m+n}(z')\,.
\ee
This is the Kac-Moody algebra associated to the conformally soft gluon sector found in~\cite{Strominger:2021lvk}.

\medskip

The worldsheet OPE also makes it easy to compute the action of these symmetry currents on hard gluons. This can also be found by taking one of the gluons in the OPE of two hard particles to be soft, but using the methods of the hard-hard case we can give an independent derivation of the soft-hard OPE. Take a soft and a hard operator
\begin{multline}
 R^{k,\msf{a}}(z_i,\bar z_i)\,\cU^{\msf{b},\veps}_{+,\Delta}(z_j,\bar z_j)\sim \frac{\im^{1-k}f^{\msf{a} \msf{b} \msf{c}}}{(1-k)!}\int\frac{j^{\msf{c}}(\sigma_j)}{\sigma_{ij}}\, \bar\delta_{k-1}(\la\lambda(\sigma_i)\,z_i\ra)\,[\mu(\sigma_i)\,\bar z_i]^{1-k}\\
 \times \bar\delta_{\Delta-1}\bigl(\la\lambda(\sigma_j)\,z_j\ra\bigr)\,\frac{(-\im\,\veps)^{1-\Delta}\,\Gamma(\Delta-1)}{[\mu(\sigma_j)\,\bar z_j]^{\Delta-1}}\,.
\end{multline}
We pick the $\sigma_{ij}^{-1}$ pole by applying \eqref{hdelta1} to the first delta function and integrating by parts, and use the fact that $\la\iota\,\lambda(\sigma_j)\ra/\la z_i\,\lambda(\sigma_j)\ra =z^{-1}_{ij}$ on the support of the second delta function, to give
\begin{multline}
 R^{k,\msf{a}}(z_i,\bar z_i)\,\cU^{\msf{b},\veps}_{+,\Delta}(z_j,\bar z_j)\sim \frac{f^{\msf{a} \msf{b} \msf{c}}}{z_{ij}}\,\frac{\im^{1-k}}{(1-k)!}\int\,j^{\msf{c}}(\sigma_j)\,\bar\delta_{\Delta+k-2}(\la\lambda(\sigma_j)\,z_j\ra)\\
 \times[\mu(\sigma_j)\,\bar z_i]^{1-k}\,\frac{(-\im\,\veps)^{1-\Delta}\,\Gamma(\Delta-1)}{[\mu(\sigma_j)\,\bar z_j]^{\Delta-1}}\,.
\end{multline}
The factors of $[\mu(\sigma_j)\,\bar z_i]^{1-k}$ now need to be replaced by derivatives in the external data $z_j,\bar z_j$ acting on a hard gluon of conformal weight $\Delta+k-1$.

To do this, note the identities
\be
\mu^{\dot1}\,\frac{\Gamma(a)}{[\mu\,\bar z_j]^{a}} = -\dbar_j\,\frac{\Gamma(a-1)}{[\mu\,\bar z_j]^{a-1}}\,,\qquad\mu^{\dot0}\frac{\Gamma(a)}{[\mu\,\bar z_j]^{a}} = (\bar z_j\dbar_j+a-1)\,\frac{\Gamma(a-1)}{[\mu\,\bar z_j]^{a-1}}\,,
\ee
where $\dbar_j\equiv\p/\p\bar z_j$ and $a\neq1$. This enables the $\mu^{\dal}$-dependent part of the OPE to be rewritten as 
\be\label{muid1}
[\mu(\sigma_j)\,\bar z_i]^{1-k}\,\frac{\Gamma(\Delta-1)}{[\mu(\sigma_j)\,\bar z_j]^{\Delta-1}} = \prod_{r=1}^{1-k}\bigl(-\bar z_{ij}\,\bar\p_j + \Delta-1-r\bigr)\,\frac{\Gamma(\Delta+k-2)}{[\mu(\sigma_j)\,\bar z_j]^{\Delta+k-2}}\,.
\ee
On the right, we recognize the $\mu^{\dal}$-dependence of a hard gluon of dimension $\Delta+k-1$. Putting it all together, we have\footnote{The $k=1$ leading soft limit case is meant to be understood as containing no derivative operators.}
\be\label{fac1}
R^{k,\msf{a}}(z_i,\bar z_i)\,\cU^{\msf{b},\veps}_{+,\Delta}(z_j,\bar z_j) \sim \frac{(-\veps)^{k-1}}{(1-k)!}\,\frac{f^{\msf{a} \msf{b} \msf{c}}}{z_{ij}} \prod_{r=1}^{1-k}\bigl(-\bar z_{ij}\,\bar\p_j + \Delta-1-r\bigr)\,\cU^{\msf{c},\veps}_{+,\Delta+k-1}(z_j,\bar z_j)\,.
\ee
For $k=1,0$, one easily verifies the actions of leading and sub-leading soft gluon symmetries.

An analogous calculation holds for the action of $R^{k,\msf{a}}$ on a negative helicity gluon. Following the methods of Section~\ref{sec:gluon+-}, we first find
\begin{multline}\label{R-}
R^{k,\msf{a}}(z_i,\bar z_i)\,\cU^{\msf{b},\veps}_{-,\Delta}(z_j,\bar z_j)\sim\frac{\im^{1-k}f^{\msf{a} \msf{b} \msf{c}}}{(1-k)!}\int\frac{j^{\msf{c}}(\sigma_j)}{\sigma_{ij}}\int\frac{\d s_i}{s_i^k}\,\frac{\d \tilde s_j}{\tilde s_j}\,\frac{\d t_j}{t_j^{2-\Delta}}\,:\bar\delta^2\biggl(z_i-s_i\lambda(\sigma_i)-\frac{\veps s_i \tilde s_j t_j z_j}{\sigma_{ij}}\biggr)\\
\times \e^{\im\veps \tilde s_jt_j\la\tilde\mu(\sigma_j)\,z_j\ra}:\,:[\mu(\sigma_i)\,\bar z_i]^{1-k}:\,:\bar\delta^2(\bar z_j-\tilde s_j\tilde\lambda(\sigma_j)):\,.
\end{multline}
Next, we compute the OPE
\begin{multline}
:[\mu(\sigma_i)\,\bar z_i]^{1-k}:\,:\bar\delta^2(\bar z_j-\tilde s_j\tilde\lambda(\sigma_j)):\\
\sim \sum_{r=0}^{1-k} {1-k\choose r}\,\frac{\im^r\,(-1)^r}{\sigma_{ij}^r}\left[\bar z_i\;\frac{\p}{\p\tilde\lambda(\sigma_j)}\right]^r\,:[\mu(\sigma_j)\,\bar z_i]^{1-k-r}\,\bar\delta^2(\bar z_j-\tilde s_j\tilde\lambda(\sigma_j)):\,.
\end{multline}
Substituting this into \eqref{R-} and then rescaling $s_i\mapsto s_i\,\sigma_{ij}$, we get
\begin{multline}
R^{k,\msf{a}}(z_i,\bar z_i)\,\cU^{\msf{b},\veps}_{-,\Delta}(z_j,\bar z_j)\sim\frac{\im^{1-k}f^{\msf{a} \msf{b} \msf{c}}}{(1-k)!}\int j^{\msf{c}}(\sigma_j)\int\frac{\d s_i}{s_i^k}\,\frac{\d \tilde s_j}{\tilde s_j}\,\frac{\d t_j}{t_j^{2-\Delta}}\,\bar\delta^2\bigl(z_i-s_i\sigma_{ij}\lambda(\sigma_i)-\veps s_i \tilde s_j t_j z_j\bigr)\\
\times \e^{\im\veps \tilde s_jt_j\la\tilde\mu(\sigma_j)\,z_j\ra}\sum_{r=0}^{1-k} {1-k\choose r}\,\frac{\im^r\,(-1)^r}{\sigma_{ij}^{k+r}}\left[\bar z_i\;\frac{\p}{\p\tilde\lambda(\sigma_j)}\right]^r\,\frac{\bar\delta^2(\bar z_j-\tilde s_j\tilde\lambda(\sigma_j))}{[\mu(\sigma_j)\,\bar z_i]^{k-1+r}}\,,
\end{multline}
with all quantities now implicitly normal-ordered.

Since $k+r\leq1$ for $r=0,\dots,1-k$, the $r=1-k$ term in the sum contains a simple pole at $\sigma_{ij}=0$ whereas the other terms in the sum are regular at $\sigma_{ij}=0$. Using this fact, we can integrate out $s_i$ and subsequently perform an integration-by-parts using the first delta function to take a residue at $\sigma_{ij}=0$. Since only the $r=1-k$ term contributes non-trivially, this dramatically simplifies the result to
\begin{multline}\label{R-1}
R^{k,\msf{a}}(z_i,\bar z_i)\,\cU^{\msf{b},\veps}_{-,\Delta}(z_j,\bar z_j)\sim\frac{(-\veps)^{k-1}}{(1-k)!}\,\frac{f^{\msf{a} \msf{b} \msf{c}}}{z_{ij}}\int j^{\msf{c}}(\sigma_j)\,\frac{(-\im\,\veps)^{2-\Delta-k}\,\Gamma(\Delta+k-2)}{\la\tilde\mu(\sigma_j)\,z_j\ra^{\Delta+k-2}}\\
\times(-1)^{1-k}\left[\bar z_i\;\frac{\p}{\p\tilde\lambda(\sigma_j)}\right]^{1-k}\,[\tilde\iota\,\tilde\lambda(\sigma_j)]^\Delta\,\bar\delta([\tilde\lambda(\sigma_j)\,\bar z_j])\,,
\end{multline}
where the remaining $\tilde s_j$ and $t_j$ integrals have been performed, and $\tilde\iota_{\dal} = (0,1)$ is a reference spinor.

At this stage, we exchange $\tilde\lambda_{\dal}(\sigma_j)$-derivatives for derivatives in the celestial positions using the identities
\be
\begin{split}
&\frac{\p}{\p\tilde\lambda_{\dot0}}\,[\tilde\iota\,\tilde\lambda]^a\,\bar\delta([\tilde\lambda\,\bar z_j]) = (\bar z_j\,\dbar_j+a)\,[\tilde\iota\,\tilde\lambda]^{a-1}\,\bar\delta([\tilde\lambda\,\bar z_j])\,,\\
&\frac{\p}{\p\tilde\lambda_{\dot1}}\,[\tilde\iota\,\tilde\lambda]^a\,\bar\delta([\tilde\lambda\,\bar z_j]) = -\dbar_j\,[\tilde\iota\,\tilde\lambda]^{a-1}\,\bar\delta([\tilde\lambda\,\bar z_j])\,,
\end{split}
\ee
where we have noted $[\tilde\iota\,\tilde\lambda] = \tilde\lambda_{\dot0}$ and $[\tilde\lambda\,\bar z_j] = \tilde\lambda_{\dot1} - \bar z_j\,\tilde\lambda_{\dot0}$. This gives
\begin{multline}\label{muid2}
(-1)^{1-k}\left[\bar z_i\;\frac{\p}{\p\tilde\lambda(\sigma_j)}\right]^{1-k}\,[\tilde\iota\,\tilde\lambda(\sigma_j)]^\Delta\,\bar\delta([\tilde\lambda(\sigma_j)\,\bar z_j]) \\
= \prod_{r=1}^{1-k}\bigl(-\bar z_{ij}\,\bar\p_j + \Delta+1-r\bigr)\,[\tilde\iota\,\tilde\lambda(\sigma_j)]^{\Delta+k-1}\,\bar\delta([\tilde\lambda(\sigma_j)\,\bar z_j])\,.
\end{multline}
Inserting this into \eqref{R-1} and recognizing a negative helicity gluon of weight $\Delta+k-1$ on the right, we find the celestial OPE
\be\label{fac2}
R^{k,\msf{a}}(z_i,\bar z_i)\,\cU^{\msf{b},\veps}_{-,\Delta}(z_j,\bar z_j) \sim \frac{(-\veps)^{k-1}}{(1-k)!}\,\frac{f^{\msf{a} \msf{b} \msf{c}}}{z_{ij}} \prod_{r=1}^{1-k}\bigl(-\bar z_{ij}\,\bar\p_j + \Delta+1-r\bigr)\,\cU^{\msf{c},\veps}_{-,\Delta+k-1}(z_j,\bar z_j)\,,
\ee
which provides the action of a positive helicity symmetry generator on a negative helicity gluon.

In summary, the OPEs in \eqref{fac1} and \eqref{fac2} can be combined into
\be\label{ggfact}
R^{k,\msf{a}}(z_i,\bar z_i)\,\cU^{\msf{b},\veps}_{J,\Delta}(z_j,\bar z_j) \sim \frac{\veps^{k-1}}{(1-k)!}\,\frac{f^{\msf{a} \msf{b} \msf{c}}}{z_{ij}} \prod_{r=1}^{1-k}\bigl(\bar z_{ij}\,\bar\p_j - 2\bar h+r\bigr)\,\cU^{\msf{c},\veps}_{J,\Delta+k-1}(z_j,\bar z_j)\,,
\ee
where $J=\pm1$ is helicity and $\bar h = (\Delta-J)/2$ is the anti-holomorphic conformal weight. This gives the result in a novel factorized form that makes contact with its ambitwistorial origins. Expanding the product of derivative operators, one equivalently finds
\begin{multline}
 R^{k,\msf{a}}(z_i,\bar z_i)\,\cU^{\msf{b},\veps}_{J,\Delta}(z_j,\bar z_j)\sim\\\frac{f^{\msf{a}\msf{b}\msf{c}}}{z_{ij}}\sum_{\ell=0}^{1-k}\frac{(-1)^{k+\ell-1}\,\veps^{k-1}}{\ell!\,(1-k-\ell)!}\frac{\Gamma(2\bar h)}{\Gamma(2\bar h+k+\ell-1)}\,\bar z_{ij}^\ell\,\bar\partial_j^\ell\,\cU^{\msf{c},\veps}_{J,\Delta+k-1}(z_j,\bar z_j)\,.
\end{multline}
This agrees with the result obtained from BCFW recursion in~\cite{Jiang:2021ovh}.


\subsection{Soft graviton symmetries}

Conformally soft graviton symmetries are associated to Poisson diffeomorphisms of twistor space or dual twistor space, lifted to ambitwistor space~\cite{Adamo:2021lrv}. They consist of two copies of the loop algebra $Lw_{1+\infty}$ of the algebra of $w_{1+\infty}$ of two-dimensional area-preserving/Poisson diffeomorphisms. Once again, we can get operator realizations of its currents from the soft expansion of vertex operators.

Recall the positive helicity conformal primary graviton vertex operator,
\be
\cV^\veps_{+,\Delta}(z,\bar z) = \int\left(\tilde\lambda^{\dal}\,\frac{\p}{\p\mu^{\dal}} + L^{\dal\dot\beta}\,\frac{\p^2}{\p\mu^{\dal}\p\mu^{\dot\beta}}\right)h_\Delta^\veps(\sigma\,|\,z,\bar z)\,,
\ee
where
\be
L^{\dal\dot\beta} = \tilde\rho^{(\dal}\rho^{\dot\beta)}
\ee
are the worldsheet currents corresponding to self-dual angular momenta, and
\be
h_\Delta^\veps(\sigma\,|\,z,\bar z) = \bar\delta_{\Delta}\bigl(\la\lambda(\sigma)\,z\ra\bigr)\,\frac{(-\im\,\veps)^{-\Delta}\,\Gamma(\Delta-2)}{[\mu(\sigma)\,\bar z]^{\Delta-2}}
\ee
is a conformal primary graviton twistor representative. Taking its residues at $\Delta=k=2,1,0,-1,\dots$ produces the soft graviton vertex operators,
\be
\begin{split}
H^k(z,\bar z) &=  \text{Res}_{\Delta=k}\,\cV^{+}_{+,\Delta}(z,\bar z)\\
&= \int\left(\tilde\lambda^{\dal}\,\frac{\p}{\p\mu^{\dal}} + L^{\dal\dot\beta}\,\frac{\p^2}{\p\mu^{\dal}\p\mu^{\dot\beta}}\right)\frac{\im^{-k}\,[\mu(\sigma)\,\bar z]^{2-k}}{(2-k)!}\,\bar\delta_{k}\bigl(\la\lambda(\sigma)\,z\ra\bigr)\\
&= \frac{1}{2\pi\im}\oint\left(\tilde\lambda^{\dal}\,\frac{\p}{\p\mu^{\dal}} + L^{\dal\dot\beta}\,\frac{\p^2}{\p\mu^{\dal}\p\mu^{\dot\beta}}\right)\frac{\im^{-k}\,[\mu(\sigma)\,\bar z]^{2-k}}{(2-k)!}\,\frac{\la\iota\,\lambda(\sigma)\ra^{k+1}}{\la\lambda(\sigma)\,z\ra}\,.
\end{split}
\ee
where we have restricted to the case of outgoing gravitons $\veps=+1$ for simplicity.

To get the currents for $w_{1+\infty}$, we relabel $k=4-2p$, set $[\mu\,\bar z] = \mu^{\dot0}+\bar z\,\mu^{\dot1}$, and expand in $\bar z$ as before:
\be
H^{4-2p}(z,\bar z) = \sum_{m=1-p}^{p-1}\frac{\bar z^{p-m-1}\,w^p_m(z)}{\Gamma(p-m)\,\Gamma(p+m)}\,.
\ee
The chiral currents $w^p_m(z)$ take the form
\be\label{wpm}
w^p_m(z) = \frac{\im^{2p}}{2\pi\im}\oint\left(\tilde\lambda^{\dal}\,\frac{\p g^p_m}{\p\mu^{\dal}} + L^{\dal\dot\beta}\,\frac{\p^2g^p_m}{\p\mu^{\dal}\p\mu^{\dot\beta}}\right)\!(\sigma)\,\frac{\la\iota\,\lambda(\sigma)\ra^{5-2p}}{\la\lambda(\sigma)\,z\ra}\,.
\ee
with the $g^p_m$ from \eqref{gpm} acting as Hamiltonians for Poisson diffeomorphism of the $\mu^{\dal}$-plane. Their OPE yields
\begin{multline}
w^p_m(z)\,w^q_n(z') \\
\sim \frac{\im^{2(p+q-2)}}{2\pi\im\,(z-z')}\oint\left(\tilde\lambda^{\dal}\,\frac{\p\{g^p_m,g^q_n\}}{\p\mu^{\dal}} + L^{\dal\dot\beta}\,\frac{\p^2\{g^p_m,g^q_n\}}{\p\mu^{\dal}\p\mu^{\dot\beta}}\right)\!(\sigma)\,\frac{\la\iota\,\lambda(\sigma)\ra^{5-2(p+q-2)}}{\la\lambda(\sigma)\,z'\ra}\,,
\end{multline}
where we have introduced the Poisson bracket on twistor space,
\be\label{gpmgqn}
\{g^p_m,g^q_n\} \equiv \veps^{\dal\dot\beta}\,\frac{\p g^p_m}{\p\mu^{\dal}}\,\frac{\p g^q_n}{\p\mu^{\dot\beta}} = 2\,\bigl(m\,(q-1)-n\,(p-1)\bigr)\,g^{p+q-2}_{m+n}\,.
\ee
Hence, we find the expected $w_{1+\infty}$ current algebra~\cite{Strominger:2021lvk}
\be
w^p_m(z)\,w^q_n(z') \sim \frac{2\,\bigl(m\,(q-1)-n\,(p-1)\bigr)}{z-z'}\,w^{p+q-2}_{m+n}(z')\,.
\ee
A similar analysis works for the negative helicity sector by parity symmetry.

We can also compute the action of a soft graviton symmetry on a soft gluon. Computing the OPE of \eqref{wpm} with \eqref{Spm} generates
\be
w^p_m(z)\,S^{q,\msf{a}}_n(z') \sim \frac{\im^{2(p+q-2)-2}}{2\pi\im\,(z-z')}\oint\frac{j^\msf{a}(\sigma)\,\{g^p_m,g^q_n\}(\sigma)}{\la\iota\,\lambda(\sigma)\ra^{2(p+q-2)-3}\,\la\lambda(\sigma)\,z'\ra}\,.
\ee
Again applying the Poisson bracket \eqref{gpmgqn}, one gets the expected celestial OPE
\be
w^p_m(z)\,S^{q,\msf{a}}_n(z') \sim \frac{2\,\bigl(m\,(q-1)-n\,(p-1)\bigr)}{z-z'}\,S^{p+q-2,\msf{a}}_{m+n}(z')\,,
\ee
demonstrating that soft gluon generators are Kac-Moody primaries of $Lw_{1+\infty}$.

\medskip

Similarly, we can compute the action of a soft graviton on a hard one. The first part of the computation is essentially the same as the hard-hard graviton OPE. After performing the integration by parts to pick the $\sigma_{ij}^{-1}$ pole and doing the leftover $t$ integral we have the simple expression
\begin{multline}\label{hsgrav1}
 H^k(z_i,\bar z_i)\,\cV^\veps_{+,\Delta}(z_j,\bar z_j)\sim
\frac{\im^{-k}}{(1-k)!}\,\frac{\bar z_{ij}}{z_{ij}}\int\bar\delta_{\Delta+k}(\la\lambda(\sigma_j)\,z_j\ra)\\
 \times\left(\tilde\lambda^{\dot\alpha}\frac{\partial}{\partial\mu^{\dot\alpha}}+\tilde\rho^{\dal}\rho^{\dot\beta}\frac{\partial^2}{\partial\mu^{\dot\alpha}\partial\mu^{\dot\beta}}\right)\!(\sigma_j)\, [\mu(\sigma_j)\,\bar z_i]^{1-k}\,\frac{(-\im\,\veps)^{-\Delta}\,\Gamma(\Delta-1)}{[\mu(\sigma_j)\,\bar z_j]^{\Delta-1}}\,.
\end{multline}
This is strictly valid for $k=1,0,-1,\dots$, as the soft graviton theorem for $k=2$ is trivial. From \eqref{hsgrav1}, we proceed in the same manner as in the soft-hard gluon-gluon OPE. The identities \eqref{muid1} give
\be
H^k(z_i,\bar z_i)\,\cV^\veps_{+,\Delta}(z_j,\bar z_j)\sim\frac{(-\veps)^{k}}{(1-k)!}\,\frac{\bar z_{ij}}{z_{ij}}\prod_{r=1}^{1-k}\bigl(-\bar z_{ij}\,\bar\p_j + \Delta-1-r\bigr)\,\cV^\veps_{+,\Delta+k}(z_j,\bar z_j)\,.
\ee
Similarly, the action on a negative helicity graviton is found to be 
\begin{multline}
 H^k(z_i,\bar z_i)\,\cV^\veps_{-,\Delta}(z_j,\bar z_j)\sim\frac{(-\veps)^{k}}{(1-k)!}\,\frac{\bar z_{ij}}{z_{ij}}\int\left(\lambda^{\alpha}\frac{\partial}{\partial\tilde\mu^{\alpha}}+\rho^{\al}\tilde\rho^{\beta}\frac{\partial^2}{\partial\tilde\mu^{\alpha}\partial\tilde\mu^{\beta}}\right)\!(\sigma_j)\\
\times\frac{(-\im\,\veps)^{2-\Delta-k}\,\Gamma(\Delta+k-2)}{\la\tilde\mu(\sigma_j)\,z_j\ra^{\Delta+k-2}}\,(-1)^{1-k}\left[\bar z_i\;\frac{\p}{\p\tilde\lambda(\sigma_j)}\right]^{1-k}\,[\tilde\iota\,\tilde\lambda(\sigma_j)]^{\Delta+2}\,\bar\delta([\tilde\lambda(\sigma_j)\,\bar z_j])\,.
\end{multline}
This is easily seen to yield
\be
H^k(z_i,\bar z_i)\,\cV^\veps_{-,\Delta}(z_j,\bar z_j)\sim\frac{(-\veps)^{k}}{(1-k)!}\,\frac{\bar z_{ij}}{z_{ij}}\prod_{r=1}^{1-k}\bigl(-\bar z_{ij}\,\bar\p_j + \Delta+3-r\bigr)\,\cV^\veps_{-,\Delta+k}(z_j,\bar z_j)\,,
\ee
having applied \eqref{muid2}.

Put together, the action of a positive helicity symmetry generator on a graviton of helicity $J=\pm2$ and weight $\bar h = (\Delta-J)/2$ is given by
\be\label{grgrfac}
H^k(z_i,\bar z_i)\,\cV^\veps_{J,\Delta}(z_j,\bar z_j)\sim-\frac{\veps^{k}}{(1-k)!}\,\frac{\bar z_{ij}}{z_{ij}}\prod_{r=1}^{1-k}\bigl(\bar z_{ij}\,\bar\p_j - 2\bar h-1+r\bigr)\,\cV^\veps_{J,\Delta+k}(z_j,\bar z_j)\,.
\ee
This is straightforwardly expanded out to produce
\be
H^k(z_i,\bar z_i)\,\cV^\veps_{J,\Delta}(z_j,\bar z_j)\sim
\frac{\bar z_{ij}}{z_{ij}}\sum_{\ell=0}^{1-k}\frac{(-1)^{k+\ell}\,\veps^{k}}{\ell!\,(1-k-\ell)!}\frac{\Gamma(2\bar h+1)}{\Gamma(2\bar h+k+\ell)}\,\bar z_{ij}^\ell\,\bar\partial_j^\ell\,\cV^\veps_{J,\Delta+k}(z_j,\bar z_j),
\ee
matching previous computations in~\cite{Himwich:2021dau,Jiang:2021ovh}.

The action of a soft graviton on a hard gluon can also be computed in a similar fashion. For a gluon of weight $\bar h = (\Delta-J)/2$, $J=\pm1$, one finds the factorized result
\be\label{sgravhglu}
H^k(z_i,\bar z_j)\,\cU^{\msf{a},\veps}_{J,\Delta}(z_j,\bar z_j)\sim-\frac{\veps^{k}}{(1-k)!}\,\frac{\bar z_{ij}}{z_{ij}}\prod_{r=1}^{1-k}\bigl(\bar z_{ij}\,\bar\p_j - 2\bar h-1+r\bigr)\,\cU^{\msf{a},\veps}_{J,\Delta+k}(z_j,\bar z_j)\,,
\ee
which can be expanded out to
\be
H^k(z_i,\bar z_i)\,\cU^{\msf{a},\veps}_{J,\Delta}(z_j,\bar z_j)\sim
\frac{\bar z_{ij}}{z_{ij}}\sum_{\ell=0}^{1-k}\frac{(-1)^{k+\ell}\,\veps^{k}}{\ell!\,(1-k-\ell)!}\frac{\Gamma(2\bar h+1)}{\Gamma(2\bar h+k+\ell)}\,\bar z_{ij}^\ell\,\bar\partial_j^\ell\,\cU^{\msf{a},\veps}_{J,\Delta+k}(z_j,\bar z_j),
\ee
again matching the action found previously in~\cite{Himwich:2021dau}


\section{Momentum eigenstates \& collinear splitting functions}
\label{Sec:MomEig}

Up to this point, our focus has been on vertex operators in the ambitwistor string corresponding to gluons and gravitons in the conformal primary basis naturally suited to CCFT. Having established that, in all cases, the worldsheet OPE between these vertex operators gives the corresponding celestial OPE in CCFT, one could ask what the worldsheet OPE corresponds to for other representations of the wavefunctions? The most natural alternative is a momentum eigenstate basis; as discussed in Section~\ref{Prelim}, the momentum space analogue of the OPE limit on the celestial sphere is the collinear limit. Thus, we expect that the worldsheet OPE between ambitwistor string vertex operators for momentum eigenstates will produce collinear splitting functions in momentum space. 

Indeed, this is easily seen to be the case; as the calculations are practically identical to (and in many cases even simpler than) those for the conformal primary basis, we only work through the gluon-gluon worldsheet OPEs here, leaving other cases as an exercise for the reader. The ambitwistor wavefunctions for positive and negative helicity (outgoing) gluons with on-shell 4-momentum $k^{\alpha\dot\alpha}=\kappa^{\alpha}\,\tilde{\kappa}^{\dot\alpha}$ are
\be\label{momeig1}
a(Z)=\int_{\C^*}\frac{\d s}{s}\,\bar{\delta}^2\!\left(\kappa-s\,\lambda\right)\,\e^{\im\,s\,[\mu\,\tilde{\kappa}]}\,, \qquad \tilde{a}(W)=\int_{\C^*}\frac{\d\tilde{s}}{\tilde{s}}\,\bar{\delta}^2\!\left(\tilde\kappa-\tilde{s}\,\tilde{\lambda}\right)\,\e^{\im\,\tilde{s}\,\la\tilde{\mu}\,\kappa\ra}\,,
\ee
which are -- as expected -- related to \eqref{cgluonsf} by a Mellin transformation. The corresponding ambitwistor string vertex operators will be denoted by
\be\label{mevo}
\begin{split}
\cU^{\msf{a}}_{+\,i}&=\int j^{\msf{a}}(\sigma_i)\,\frac{\d s_i}{s_i}\,\bar{\delta}^2\!\left(\kappa_i-s_i\,\lambda(\sigma_i)\right)\,\e^{\im\,s_i\,[\mu(\sigma_i)\,i]}\,, \\
\cU^{\msf{a}}_{-\,i}&=\int j^{\msf{a}}(\sigma_i)\,\frac{\d\tilde{s}_i}{\tilde{s}_i}\,\bar{\delta}^2\!\left(\tilde\kappa_i-\tilde{s}_i\,\tilde{\lambda}(\sigma_i)\right)\,\e^{\im\,\tilde{s}_i\,\la\tilde{\mu}(\sigma_i)\,i\ra}\,,
\end{split}
\ee
respectively.

For gluons of the same helicity, the worldsheet OPE gives
\begin{multline}\label{mesh1}
\cU^{\msf{a}}_{+\,i}\,\cU^{\msf{b}}_{+\,j}\sim\int\d\sigma_i\,\frac{f^{\msf{abc}}\,j^{\msf{c}}(\sigma_j)}{\sigma_{ij}}\,\frac{\d s_i}{s_i}\,\frac{\d s_j}{s_j}\,\e^{\im\,s_i\,[\mu(\sigma_i)\,i]+\im\,s_j\,[\mu(\sigma_j)\,j]} \\
\bar{\delta}^2\!\left(\kappa_i-s_i\,\lambda(\sigma_i)\right)\,\bar{\delta}^2\!\left(\kappa_j-s_j\,\lambda(\sigma_j)\right)\,.
\end{multline}
Inspecting the holomorphic delta functions, it follows that $\sigma_{ij}\to0$ on the worldsheet now corresponds to $\la i\,j\ra\to0$ in momentum space, consistent with the (holomorphic) collinear limit. After performing the $s_i$ integral against one of the holomorphic delta functions, we integrate-by-parts to localize on $\sigma_{ij}=0$, after which the OPE is reduced to
\begin{multline}\label{mesh2}
\cU^{\msf{a}}_{+\,i}\,\cU^{\msf{b}}_{+\,j}\sim f^{\msf{abc}}\,\frac{\la\xi\,j\ra}{\la\xi\,i\ra\,\la i\,j\ra}\,\int j^{\msf{c}}(\sigma_j)\,\frac{\d s_j}{s_j}\,\bar{\delta}^2\!\left(\kappa_j-s_j\,\lambda(\sigma_j)\right) \\
 \exp\left[\frac{\im\,s_j}{\la\xi\,j\ra}\left(\la\xi\,i\ra\,[\mu(\sigma_j)\,i]+\la\xi\,j\ra\,[\mu(\sigma_j)\,j]\right)\right]\,,
\end{multline}
where $\xi^{\alpha}$ is an arbitrary reference spinor used to perform the $s_i$ integral. Now, the collinear limit can be parametrized by
\be\label{coll}
k_i+k_j=p+\epsilon^2\,\xi^\alpha\,\tilde{\xi}^{\dot\alpha}\,,
\ee
where $p^{\alpha\dot\alpha}=p^{\alpha}\,\tilde{p}^{\dot\alpha}$ is the collinear null momentum and $\epsilon$ is a parameter controlling the collinear limit (i.e., $\epsilon\to0$). After re-scaling $s_j\mapsto s_j \frac{\la\xi\,j\ra}{\la\xi\,p\ra}$, the OPE can be written in terms of the collinear momentum as
\be\label{mesh3}
\cU^{\msf{a}}_{+\,i}\,\cU^{\msf{b}}_{+\,j}\sim \frac{f^{\msf{abc}}}{\la i\,j\ra}\,\frac{\la\xi\,p\ra^2}{\la\xi\,i\ra\,\la\xi\,j\ra}\,\cU^{\msf{c}}_{+\,p}+O(\epsilon)\,.
\ee
The worldsheet OPE coefficient is precisely the collinear splitting function of two positive helicity gluons~\cite{Altarelli:1977zs,Birthwright:2005ak}:
\be\label{spliteq}
\mbox{Split}(i_{+}^{\msf{a}}, j^{\msf{b}}_{+}\rightarrow p_{+}^{\msf{c}})=\frac{f^{\msf{abc}}}{\la i\,j\ra}\,\frac{\la\xi\,p\ra^2}{\la\xi\,i\ra\,\la\xi\,j\ra}\,,
\ee
written in spinor helicity variables. Note that changing the reference spinor $\xi^{\alpha}$ alters the splitting function but does not change the singular part of the splitting function, so the choice of $\xi^{\alpha}$ is indeed arbitrary in the collinear limit.

For gluons of opposite helicity, the worldsheet OPE produces holomorphic delta functions which are singular as $\sigma_{ij}\to0$:
\begin{multline}\label{meoh1}
\cU^{\msf{a}}_{+\,i}\,\cU^{\msf{b}}_{-\,j} \sim\int \d\sigma_i\,\frac{f^{\msf{abc}}\,j^{c}(\sigma_j)}{\sigma_{ij}}\,\frac{\d s_i}{s_i}\,\frac{\d\tilde{s}_j}{\tilde{s}_j}\,\e^{\im\,s_i\,[\mu(\sigma_i)\,i]+\im\,\tilde{s}_j\,\la\tilde{\mu}(\sigma_j)\,j\ra}\\
\bar{\delta}^{2}\!\left(z_i-s_{i}\,\lambda(\sigma_i)-\frac{s_i\,\tilde{s}_j\,\kappa_j}{\sigma_{ij}}\right)\,\bar{\delta}^{2}\!\left(\tilde{\kappa}_j-\tilde{s}_j\,\tilde{\lambda}(\sigma_j)+\frac{s_i\,\tilde{s}_j\,\tilde{\kappa}_i}{\sigma_{ij}}\right)\,.
\end{multline}
However, if one requires that the arguments of the delta functions remain non-singular as $\sigma_{ij}\to0$ it is clear that they force this limit to correspond with the (holomorphic and anti-holomorphic) collinear limit. To proceed, we follow the same strategy as in Section~\ref{Sec:Mixed}: evaluate on two reparametrized coordinate patches $s_i\mapsto s_i\,\sigma_{ij}$ and $\tilde{s}_j\mapsto\tilde{s}_j\,\sigma_{ij}$. In each patch the remainder of the calculation is actually much simpler than for the conformal primary basis (as there are fewer additional re-scalings required to evaluate the remaining integrals); the final result is
\be\label{meoh2}
\cU^{\msf{a}}_{+\,i}\,\cU^{\msf{b}}_{-\,j} \sim \frac{f^{\msf{abc}}}{\la i\,j\ra}\,\frac{\la\xi\,j\ra}{\la\xi\,i\ra}\,\cU^{\msf{c}}_{-\,p}+ \frac{f^{\msf{abc}}}{[j\,i]}\,\frac{[\tilde{\xi}\,i]}{[\tilde{\xi}\,j]}\,\cU^{\msf{c}}_{+\,p}+O(\epsilon)\,,
\ee
which produces the correct collinear splitting functions for both holomorphic and anti-holomorphic collinear limits~\cite{Altarelli:1977zs,Birthwright:2005ak}. 

The same calculations for graviton-graviton and gluon-graviton OPEs with momentum eigenstates proceed along similar lines. The details of the calculations recapitulate much of what has already been covered for the conformal primary basis, so we do not include them here. The outcome is, of course, the same: the worldsheet OPE produces the correct collinear splitting functions for (tree-level) general relativity~\cite{Bern:1998sv,White:2011yy,Akhoury:2011kq} and EYM~\cite{Stieberger:2016lng,Pate:2019lpp}.

\medskip

The collinear splitting functions obtained in a momentum eigenstate basis seem fairly compact when compared against the infinite towers of $\SL(2,\R)$ descendant contributions to the celestial OPE obtained in a conformal primary basis. Since the collinear limit in momentum space is linked with the OPE limit on the celestial sphere by a Mellin transformation, one expects that there should be some avatar of the descendants visible even in momentum space. This is indeed true; the required structure emerges after explicitly parametrizing the on-shell momenta in terms of the celestial sphere; the spinorial version of \eqref{csparam} is
\be\label{csspinor}
k_i^{\alpha\dot\alpha}=\omega_i\,z_i^{\alpha}\,\bar{z}_i^{\dot\alpha}\,, \qquad \kappa_i^{\alpha}=\sqrt{\omega_i}\,z_i^{\alpha}\,, \quad \tilde{\kappa}_{i}^{\dot\alpha}=\sqrt{\omega_i}\,\bar{z}_i^{\dot\alpha}\,.
\ee
Feeding this into the same-helicity result \eqref{mesh3}, for instance, and expanding the effective momentum $p=k_i+k_j$ in $z_{ij}$ immediately enables further refinement by expanding the resulting collinear gluon vertex operator and setting $\xi^{\alpha}=\iota^{\alpha}$:
\be\label{mesh4}
\cU^{\msf{a}}_{+\,i}\,\cU^{\msf{b}}_{+\,j}\sim \frac{f^{\msf{abc}}}{z_{ij}}\,\frac{\omega_p}{\omega_i\,\omega_j}\,\exp\left(\frac{\omega_i}{\omega_p}\,\bar{z}_{ij}\,\dbar\right)\,\cU^{\msf{c}}_{+}(\omega_p,z_j,\bar{z})\Bigr|_{\bar z=\bar z_j}\,,
\ee
where the $\dbar = \p/\p\bar z$ derivatives do not act on factors of $\bar z_{ij}$ as they are evaluated before substituting $\bar z=\bar z_j$. In other words, the soft expansion of the collinear gluon in $\omega_i$ exponentiates \cite{He:2014bga,Lipstein:2015rxa,Guevara:2019ypd}; this does not alter the holomorphic collinear singularity ($z_{ij}\to0$ in these variables) but does introduce an infinite number of $\bar{z}_{ij}$ corrections with powers of $\omega_i$. It is precisely this exponentiated soft factor that produces the tower $\SL(2,\R)$ descendant contributions to the celestial OPE after a Mellin transformation.  

\medskip

In the conformal primary basis, we derived several chiral towers of incoming/outgoing celestial OPE coefficients that had not been explicitly computed before. Momentum space provides a straightforward check on these results, where they can rederived by Mellin transforming collinear expansions (in line with previous approaches~\cite{Strominger:2021lvk,Himwich:2021dau}). For example, consider two colliear positive helicity gluons; following \eqref{spliteq}, the collinear expansion of the momentum space color-stripped gluon amplitude reads
\be
A(k_1,k_2,k_3,\cdots,k_n) \sim \frac{1}{z_{12}}\,\frac{(\veps_1\,\omega_1+\veps_2\,\omega_2)^2}{\omega_1\,\omega_2}\,A(p,k_3,\cdots,k_n)\,,
\ee
using the momentum parametrization \eqref{csspinor}, with $p$ the effective momentum $p=k_1+k_2$. Expanding $p$ in $z_{12}$ gives:
\be
\begin{split}
p^{\al\dal} &= \veps_1\,\omega_1\,z_1^\al\,\bar z_1^{\dal} + \veps_2\,\omega_2\,z_2^\al\,\bar z_2^{\dal} = (\veps_1\,\omega_1\,\bar z_1^{\dal} + \veps_2\,\omega_2\,\bar z_2^{\dal})\,z_1^\al + O(z_{12})\\
&= (\veps_1\,\omega_1+\veps_2\,\omega_2)\,z_1^{\al}\,\left(\bar z_2^{\dal} + \frac{\veps_1\,\omega_1\,\bar z_{12}\,\tilde\iota^{\dal}}{\veps_1\,\omega_1+\veps_2\,\omega_2}\right) + O(z_{12})\,,
\end{split}
\ee
where $\tilde\iota_{\dal}\equiv(0,1)$. Using this to Taylor expand $A(p,k_3,\cdots,k_n)$ in $z_{12}$, $\bar z_{12}$ and only keeping terms that are singular in $z_{12}$ yields
\begin{multline}
A(k_1,k_2,k_3,\cdots,k_n) \\
\sim \frac{1}{z_{12}}\,\frac{(\veps_1\omega_1+\veps_2\omega_2)^2}{\omega_1\,\omega_2}\sum_{m=0}^\infty\frac{1}{n!}\left( \frac{\veps_1\omega_1}{\veps_1\omega_1+\veps_2\omega_2}\right)^n\,\bar z_{12}^n\,\bar\p_2^nA(k_2,k_3,\cdots,k_n) + O(z_{12}^0)\,.
\end{multline}
Mellin transforming and dressing with color factors, this can be easily checked to yield our general result \eqref{shgl7} for the celestial OPE (with indices $i=1$, $j=2$) for generic incoming-outgoing configurations. Similar computations apply to gravitons and the mixed-helicity cases.


\section{Discussion}

Ambitwistor strings provide a worldsheet description of gluon and graviton amplitudes in flat space. In the first instance, one expects that celestial CFT could act as a holographic dual to ambitwistor strings. In this work, we took the first steps in developing a map between the worldsheet OPE of ambitwistor string vertex operators and the operator algebra of celestial CFT. We saw that the worldsheet OPE provides a dynamical way to generate the celestial OPE of boost eigenstates without actually having to compute the associated amplitudes. Since our understanding of holography in other space-times is inexorably tied to string theory, we anticipate that looking for further connections between celestial holography and twistor/ambitwistor strings would prove highly fruitful.

Looking back, it is useful to compare our method with previous approaches to generating celestial OPE coefficients. In previous work, one first computed collinear splitting functions like \eqref{spliteq} using BCFW recursion relations and general factorization properties of on-shell amplitudes. Subsequently, these were Mellin transformed to obtain expansions of celestial amplitudes in the `OPE limit'. It was then conjectured that these come from the OPE of a putative celestial CFT. As a highlight of our current approach, we have provided at least one concrete construction of a 2-dimensional CFT whose operators realize various celestial OPEs through genuine operator products. This bolsters the expectation that there exists some algebra of boundary celestial operators that is isomorphic to the algebra of vertex operators coupling to ambitwistor strings.

While we have seen that the chiral-singular part of celestial OPEs can be determined dynamically using ambitwistor strings, it would be interesting to see if \emph{all} descendant contributions (including regular terms, such as those computed in~\cite{Banerjee:2020kaa,Banerjee:2020zlg,Ebert:2020nqf,Banerjee:2020vnt,Banerjee:2021dlm}) are captured by the worldsheet OPE. Clearly, regular terms are not captured by the raw worldsheet calculations we have done here, since these localize on the boundary of the moduli space and hence the singular portion of the OPE. To capture the full descendant tower one must presumably account for the OPE inside of a specific worldsheet correlation function, taking into account the full worldsheet path integral.


\acknowledgments

We would like to thank Yvonne Geyer for conversations and sharing notes on 4d ambitwistor strings, and Andrew Strominger and Tomasz Taylor for discussions. TA is supported by a Royal Society University Research Fellowship and by the Leverhulme Trust (RPG-2020-386). WB is supported by a Royal Society PhD Studentship. EC is supported by the Frankel-Goldfield-Valani Research Fund. AS is supported by a Mathematical Institute Studentship, Oxford and by the ERC grant GALOP ID: 724638.

\bibliographystyle{JHEP}
\bibliography{cope}

\end{document}